\documentclass[12pt,a4paper]{article}
\usepackage{amsmath}
\usepackage{amssymb}
\usepackage{hyperref}
\usepackage{setspace}
\usepackage{graphicx} 
\usepackage{cite}

\begin{document}

\date{}
\title{{\Large \textbf{Hydrodynamics of cold holographic matter}}}
\author{{\normalsize Richard A. Davison and Andrei Parnachev} \vspace{0.5cm} \\ {\normalsize \textit{Instituut-Lorentz for Theoretical Physics}} \\{\normalsize\textit{Niels Bohrweg 2, Leiden NL-2333 CA, The Netherlands}} \vspace{0.1cm} \\ {\normalsize davison@lorentz.leidenuniv.nl, parnachev@lorentz.leidenuniv.nl}}

\maketitle

\thispagestyle{empty}

\onehalfspacing
\begin{abstract}

{\normalsize We show that at any temperature, the low-energy (with respect to the chemical potential) collective excitations of the transverse components of the energy-momentum tensor and the global U(1) current in the field theory dual to the planar RN-AdS$_4$ black hole are simply those of hydrodynamics. That is, hydrodynamics is applicable even at energy scales much greater than the temperature. It is applicable even at zero temperature. Specifically, we find that there is always a diffusion mode with diffusion constant proportional to the ratio of entropy density to energy density. At low temperatures, the leading order momentum and temperature dependences of the dispersion relation of this mode are controlled by the dimension of an operator in the thermal CFT$_1$ dual to the near-horizon Schwarzschild-AdS$_2$ geometry.}

\end{abstract}

\clearpage
\tableofcontents

\section{Introduction}

Over the last decade, it has become clear that gauge/gravity duality \cite{Maldacena:1997re,Gubser:1998bc,Witten:1998qj} is a very useful tool for studying the hydrodynamics of strongly-coupled field theories (see \cite{Son:2007vk,CasalderreySolana:2011us,Hubeny:2011hd} for some reviews of this field). The theory of hydrodynamics \cite{LandauLifshitz} is formulated as a derivative expansion in which the macroscopic quantities describing the state of a system close to equilibrium are assumed to be slowly-varying with respect to a length scale $l_\text{mfp}$, often taken to be the mean free path between thermal collisions. Typically, increasing the temperature $T$ will reduce $l_\text{mfp}$ and thus increase the accuracy of hydrodynamics. A recent introduction to this topic can be found in \cite{Kovtun:2012rj}.

The theory of hydrodynamics predicts that for a conformal theory in $D$ spacetime dimensions deformed to non-zero $T$ and with a chemical potential $\mu$ for a conserved U(1) charge, the two-point functions of the energy-momentum tensor $T^{\mu\nu}$ and of the $U(1)$ current $J^\mu$ exhibit three kinds of gapless excitation: a transverse diffusion mode with dispersion relation (where $\omega$ is the frequency of the excitation and $q$ its momentum)
\begin{equation}
\label{eq:ExpectedHydroDiffusionResult}
\omega=-i\frac{\eta}{\epsilon+P}q^2+O\left(q^4\right),
\end{equation}
a longitudinal charge diffusion mode, and a longitudinal sound mode with dispersion relation
\begin{equation}
\label{eq:ExpectedHydroSoundResult}
\omega=\pm\sqrt{\frac{dP}{d\epsilon}}q-i\frac{\left(D-2\right)}{\left(D-1\right)}\frac{\eta}{\left(\epsilon+P\right)}q^2+O\left(q^3\right),
\end{equation}
where $\epsilon$ is the energy density and $P$ the pressure of the field theory (as the underlying theory is conformal, these are related via $\epsilon=\left(D-1\right)P$). The shear viscosity $\eta$ of the theory, which controls the attenuation of the sound and transverse diffusion modes, can be independently determined via the Kubo formula
\begin{equation}
\label{eq:KuboFormulaForViscosity}
\eta\equiv-\lim_{\omega\rightarrow0}\frac{1}{\omega}\text{Im}G^R_{T^{xy}T^{xy}}\left(\omega,q=0\right),
\end{equation}
where $G^R_{\mathcal{O}_i\mathcal{O}_j}$ denotes the retarded Greens function of a pair of operators.

Starting with \cite{Policastro:2002se,Policastro:2002tn,Herzog:2002fn,Herzog:2003ke}, the existence of these modes has been established in a wide variety of holographic theories in the limit $\omega,q\ll T$, which is analogous to the limit described above where the thermal collision length is much shorter than the wavelength of the perturbation. Remarkably, it has been shown that for a very large class of holographic theories the viscosity is given by \cite{Policastro:2001yc,Kovtun:2004de,Buchel:2004qq,Iqbal:2008by}
\begin{equation}
\label{eq:IntroEtaOverS}
\eta=\frac{s}{4\pi},
\end{equation}
where $s$ is the entropy density of the field theory. This result is significant as the ratio $\eta/s$ is similar to that measured for the quark-gluon plasma, and significantly smaller from that predicted by, for example, perturbative QCD. One obvious lesson from this work is that the transport properties of strongly-coupled theories can be significantly different from those of theories based on the existence of long-lived quasiparticles. The applicability of hydrodynamics to holographic theories can be checked to higher orders in the derivative expansion and in the amplitude of fluctuations using the fluid/gravity correspondence \cite{Bhattacharyya:2008jc,Banerjee:2008th,Rangamani:2009xk}.

A question which has had comparatively little study in the gauge/gravity duality literature is what happens when the mean free path between thermal collisions becomes very long -- when $\omega,q\gg T$, or in the extreme limit when $T=0$. One could argue that hydrodynamics -- a derivative expansion valid on length scales much longer than $l_\text{mfp}$ -- should be valid even at $T=0$ provided that we consider perturbations with $\omega,q\ll\mu$. In other words, that at $T=0$ hydrodynamics will continue to be valid since a large $\mu$ will result in a small mean free path $l_\text{mfp}\sim1/\mu$ and thus a well-controlled derivative expansion. This viewpoint was emphasised in \cite{Bhattacharyya:2007vs}: assuming that $\eta/s=1/4\pi$ at $T=0$, one can define $l_\text{mfp}$ to be equal to the diffusion constant in (\ref{eq:ExpectedHydroDiffusionResult}) and show that it is finite at $T=0$ for a certain class of large, charged black holes in global AdS spaces.

However, this is at odds with our intuition that  turning on a large $T$ or a large $\mu$ are qualitatively different deformations of a field theory. Moreover, in the best-understood quasparticle-based theory of this kind -- Fermi liquid theory -- the hydrodynamics outlined above is not valid at sufficiently low temperatures.\footnote{It breaks down when $\omega\gtrsim l_\text{mfp}^{-1}\sim T^2/\mu$.} For example, the speed of sound at zero temperature is not equal to its hydrodynamic value $\sqrt{dP/d\epsilon}$ and the decay rate of sound $-\text{Im}\left(\omega\right)\propto q^2+T^2$ has an important, $q$-independent contribution \cite{AbrikosovKhalatnikov,Pethick,PinesNozieres,BaymPethick}. Furthermore, it has been shown in various holographic probe D-brane theories that hydrodynamics is not an accurate description at sufficiently low temperatures, as the longitudinal charge diffusion mode expected from the hydrodynamics of a conserved $U(1)$ current is replaced by propagating sound modes \cite{Karch:2008fa,Kulaxizi:2008kv,Kulaxizi:2008jx,Hung:2009qk,HoyosBadajoz:2010kd,Lee:2010ez,Bergman:2011rf,Ammon:2011hz,Davison:2011ek,Goykhman:2012vy,Brattan:2012nb,Jokela:2012se}.

There have been a few studies of the low-energy bosonic excitations outside of the usual hydrodynamic limit $\omega,q\ll T$ for non-probe holographic theories with a non-zero $\mu$, including \cite{Edalati:2010hk,Edalati:2010pn,Brattan:2010pq,Davison:2011uk,Faulkner:2012gt}. The main technical difference of these calculations, compared to those in the usual hydrodynamic limit, is that as $T\rightarrow0$ the location of the black hole horizon becomes an irregular singular point of the equations of motion for the excitations of the bulk fields. 

Additionally, it has been shown that if one defines the `viscosity' at $T=0$ formally using the Kubo formula (\ref{eq:KuboFormulaForViscosity}), then for a small subset of holographic theories (which all have $s\ne0$ at $T=0$) the relation $\eta=s/4\pi$ continues to hold even at $T=0$ \cite{Edalati:2009bi,Imeroni:2009cs,Paulos:2009yk,Cai:2009zn,Chakrabarti:2009ht}. This subset includes the planar AdS-Reissner-Nordstr\"{o}m black hole. We stress here that these examples are \textit{not} covered by the general proof of \cite{Iqbal:2008by} because of the different horizon structure in the $T=0$ limit mentioned above. Physically, the two situations are completely distinct as the $\omega\rightarrow0$ limit in the definition of $\eta$ corresponds to $\omega\ll T,\mu$ at any non-zero $T$, and $T\ll\omega\ll\mu$ when $T=0$. Henceforth we will abuse notation and denote both the usual viscosity, and its zero temperature analogue, by the name `viscosity' and the symbol $\eta$ since they are both equal to $s/4\pi$ for the field theory that we investigate.

In this paper, we study the retarded Greens functions of the transverse components of $T^{\mu\nu}$ and $J^\mu$ in the field theory dual to the planar AdS-Reissner-Nordstr\"{o}m black hole in (3+1)-dimensions (RN-AdS$_4$). We analytically compute these Greens functions at leading order in small $\omega,q$ under the assumption that these quantities are small compared to \textit{either} $T$ or $\mu$ (or both). When $\omega,q\ll T$, we recover the expected hydrodynamic diffusion mode with a dispersion relation given by (\ref{eq:ExpectedHydroDiffusionResult}) and (\ref{eq:IntroEtaOverS}), in agreement with the results of \cite{Saremi:2006ep,Ge:2010yc}. More interestingly, we show that this diffusion mode exists with a dispersion relation given by (\ref{eq:ExpectedHydroDiffusionResult}) and $(\ref{eq:IntroEtaOverS})$ even when $\omega\gtrsim T$. In fact, even in the extremal limit $T=0$, we find that there exists a diffusive mode described by the hydrodynamic results (\ref{eq:ExpectedHydroDiffusionResult}) and (\ref{eq:IntroEtaOverS}). To our knowledge, this is the first analytic calculation of the bosonic excitations outside of the usual hydrodynamic limit $\omega\ll T$ in a holographic theory at non-zero density with a dynamical metric.

Outside of the usual hydrodynamic range $\omega\ll T$, our analytic calculation of the dispersion relation of the diffusion mode relies on the existence of a (Schwarzschild-)AdS$_2$ near-horizon geometry. In the spirit of the semi-holographic work of \cite{Iqbal:2008by,Faulkner:2009wj,Faulkner:2010tq,Nickel:2010pr}, we find that whereas the existence of low-energy modes (in our case diffusion) is a property of the full spacetime geometry, the decay rate of these modes is controlled by the near-horizon geometry. In our case, the leading power of the momentum in the dispersion relation $\omega\sim-iq^\alpha$ is controlled by the dimension of the operator dual to the gauge-invariant combination of the metric fluctuations in the CFT$_1$ dual to the near-horizon AdS$_2$ geometry. Formally, we find that the dispersion relation of the diffusion mode at $T=0$ can be written
\begin{equation}
\frac{\omega^2}{\mathcal{G}^\text{IR}_1\left(\omega\right)}\sim-q^2,
\end{equation}
where $\mathcal{G}^\text{IR}_1\left(\omega\right)$ is the retarded Greens function of a scalar operator of conformal dimension $1$ in the CFT$_1$ dual to the near-horizon AdS$_2$ region of the geometry. As this CFT$_1$ correlator is proportional to $i\omega$, we obtain $\omega\sim -iq^2$. Note that unlike in the fermionic case of \cite{Faulkner:2009wj}, the dimension of this operator is fixed. At non-zero temperatures $T\lesssim\omega\ll\mu$, $\mathcal{G}^\text{IR}_1\left(\omega\right)$ is replaced by $\mathcal{G}^\text{IR}_{1,T}\left(\omega\right)$, the thermal Greens function of the CFT$_1$ operator, in the above equation. This ensures that the dispersion relation $\omega\left(q\right)$ retains the same $\omega\sim-iq^2$ form (but with a $T$-dependent diffusion constant). In particular, it does not receive any corrections which are dependent upon $T$ but independent of $q$ (at lowest order).

The $T=0$ Greens functions of the transverse operators of this theory were analytically studied previously in \cite{Edalati:2009bi,Edalati:2010hk}, and our results are complementary to theirs. We use different gauge-invariant fields than in their approach, which allow us to determine the leading-order momentum dependence of these fields far from the horizon. As just indicated, it is the knowledge of this far-from-horizon solution which allows us to demonstrate the existence of a diffusion mode. On the other hand, the fields used in \cite{Edalati:2009bi,Edalati:2010hk} allow one to determine the dependence of these fields near the horizon to a higher order in $\omega$ and $q$ than in our approach. This means that the scaling of the spectral functions in the limit $\omega\rightarrow0$ is captured more accurately by the results of \cite{Edalati:2009bi,Edalati:2010hk} (although at leading order, our results agree).

The remainder of this paper is structured as follows. In section \ref{sec:BackgroundSolutionAndFluctuationsEoMs} we review the thermodynamic properties of the planar RN-AdS$_4$ solution, and manipulate the equations of motion (and on-shell action) of the fluctuations of the transverse fields into a form in which we can solve them in the appropriate limits. In section \ref{sec:HydroCalculation} we calculate the Greens functions of the dual operators of these transverse fields in the limit $\omega\ll T$. This is an instructive warmup for sections \ref{sec:ZeroTCalculation} and \ref{sec:NonHydroNonZeroTCalculation}, in which the same general procedure is used to calculate the Greens functions in the $T=0$ (with $\omega\ll\mu$) and $T\lesssim\omega\ll\mu$ limits respectively. In section \ref{sec:NumericalResults} we present results of the numerical computation of the poles of the Greens functions, and the spectral functions, at both zero and non-zero temperatures. These confirm the accuracy of our analytic results. Finally in section \ref{sec:Discussion} we conclude with a discussion of some possible future research directions.

\section{The planar RN-AdS$_4$ solution and its linearised excitations}
\label{sec:BackgroundSolutionAndFluctuationsEoMs}
\subsection{The equilibrium solution}
The planar AdS$_4$ Reissner-Nordstr\"{o}m spacetime that we study is a solution to the Einstein-Maxwell theory with a cosmological constant. The action of this theory is
\begin{equation}
\label{eq:EinsteinMaxwellequation}
S=\frac{1}{2\kappa_4^2}\int_\mathcal{M} d^4x\sqrt{-g}\left(\mathcal{R}+\frac{6}{L^2}-L^2F_{\mu\nu}F^{\mu\nu}\right)+\frac{1}{\kappa_4^2}\int_{\partial\mathcal{M}}d^3x\sqrt{\left|h\right|}\;\mathcal{K}+\text{counterterms},
\end{equation}
and the planar RN-AdS$_4$ solution can be written
\begin{equation}
\label{eq:RNAdS4solution}
\begin{aligned}
ds^2&=-\frac{r^2f(r)}{L^2}dt^2+\frac{r^2}{L^2}\left(dx^2+dy^2\right)+\frac{L^2}{r^2f(r)}dr^2, \;\;\;\;\;\;\; A_t\left(r\right)=\frac{Qr_0}{L^2}\left(1-\frac{r_0}{r}\right),\\
f(r)&=1-(1+Q^2)\frac{r_0^3}{r^3}+Q^2\frac{r_0^4}{r^4},
\end{aligned}
\end{equation}
where the boundary of the spacetime is at $r\rightarrow\infty$, the planar outer horizon is at $r=r_0$ and the U(1) field strength is given by $F=dA$. The term involving the extrinsic curvature $\mathcal{K}$ is the Gibbons-Hawking term, and we have not written explicitly the counterterms required to remove the divergences from the on-shell action. These counterterms, which can be found for example in \cite{Edalati:2010hk}, contribute only contact terms to the dual Greens functions and so will not be required for our purposes.

This solution can be consistently embedded into 11-dimensional supergravity, where the Einstein-Maxwell theory given by equation (\ref{eq:EinsteinMaxwellequation}) arises as a universal sector in the gravitational dual of any (2+1)-dimensional SCFT with $N=2$ supersymmetry \cite{Gauntlett:2007ma}. In the simplest embedding, this solution is dual to the state of the low-energy (2+1)-dimensional field theory on a stack of M2-branes in flat space with a non-zero density of diagonal U(1) R-charge \cite{Chamblin:1999tk,Cvetic:1999xp}. 

The RN-AdS$_4$ solution is the thermodynamically-preferred solution to the theory defined by the action (\ref{eq:EinsteinMaxwellequation}) at all temperatures. However, by studying more general truncations of the supergravity action (i.e.~by including the effects of more operators in the dual CFT), it is found that this solution is highly susceptible to instabilities due to the condensation of both charged and neutral scalar fields \cite{Denef:2009tp,Gauntlett:2009dn,Gauntlett:2009bh,Donos:2011ut}. For this reason, we find it unlikely that the uplifts of the solution (\ref{eq:EinsteinMaxwellequation}) are thermodynamically-preferred phases of the full 11-dimensional supergravity action (and hence of the CFTs dual to this) when $T=0$. 

The solution has one dimensionless parameter $Q$, which determines the ratio of the temperature of the field theory state $T$ to the chemical potential (with respect to the diagonal U(1) R-charge) $\mu$
\begin{equation}
\label{eq:DefinitionofBackgoundTemp}
T=\frac{\left(3-Q^2\right)r_0}{4\pi L^2},\;\;\;\;\;\;\;\;\;\;\;\;\;\;\;\mu=\frac{Qr_0}{L^2},\;\;\;\;\;\;\;\;\;\;\;\;\;\;\;\frac{T}{\mu}=\frac{3-Q^2}{4\pi Q}.
\end{equation}
This can be inverted to give $Q$ in terms of $T/\mu$
\begin{equation}
\label{eq:QasfunctionofTandMu}
Q=\sqrt{3+4\pi^2\frac{T^2}{\mu^2}}-2\pi\frac{T}{\mu}.
\end{equation}
Q can take values between $0$ (corresponding to $\mu=0$) and $\sqrt{3}$ (corresponding to $T=0$). 

The thermodynamic properties of the dual field theory state are controlled by the dimensionless parameter $Q$ (or equivalently by $T/\mu$). The energy density $\epsilon$, pressure $P$, entropy density $s$ and charge density $\rho$ are given by \cite{Chamblin:1999tk}
\begin{equation}
\epsilon=\frac{\left(1+Q^2\right)r_0^3}{\kappa_4^2L^4},\;\;\;\;\;\;\;\;\;\;P=\frac{\epsilon}{2},\;\;\;\;\;\;\;\;\;\;s=\frac{2\pi r_0^2}{\kappa_4^2L^2},\;\;\;\;\;\;\;\;\;\;\rho=\frac{2Qr_0^2}{\kappa_4^2L^2}.
\end{equation}
When $T=0$, this state has the unusual property that its entropy density is non-zero.

\subsection{Linearised fluctuations around equilibrium}

The two-point functions of field theory operators can be computed by studying the linearised fluctuations of their dual fields in the gravitational theory. Without loss of generality, we choose the momentum of these fluctuations to flow along the $x$-direction and excite the metric -- which is dual to the energy-momentum tensor $T^{\mu\nu}$ of the field theory -- and the gauge field -- which is dual to the diagonal U(1) R-current $J^\mu$ -- around their values on the planar RN-AdS$_4$ solution as follows
\begin{equation}
\begin{aligned}
g_{\mu\nu}\left(r\right)&\rightarrow g_{\mu\nu}\left(r\right)+\int\frac{d\omega dq}{\left(2\pi\right)^2}e^{-i\omega t+iqx}h_{\mu\nu}\left(r,\omega,q\right),\\
A_\mu\left(r\right)&\rightarrow A_\mu\left(r\right)+\int\frac{d\omega dq}{\left(2\pi\right)^2}e^{-i\omega t+iqx}a_\mu\left(r,\omega,q\right).
\end{aligned}
\end{equation}

At linear order, the fluctuations of fields transverse to the momentum flow -- $h_{yt}, h_{xy}, h_{ry}$ and $a_y$ -- decouple from the rest, due to the $y\rightarrow-y$ symmetry of the theory and of the planar RN-AdS$_4$ solution. In this paper we are interested only in the fluctuations of these `transverse fields'. By varying the action (\ref{eq:EinsteinMaxwellequation}) with respect to $h_{yt}, h_{xy}, h_{ry}$ and $a_y$ respectively, we obtain the following set of coupled equations of motion
\begin{subequations}
\begin{align}
& \frac{d}{dr}\left[r^4{h^y_t}'+\frac{i\omega L^4}{f}h^r_y+4L^4r^2A_t'a_y\right]-\frac{L^4q}{f}\left(qh^y_t+\omega h^x_y\right)=0,\label{eq:HytEoM}\\
& \frac{d}{dr}\left[r^4f{h^x_y}'-iqL^4h^r_y\right]+\frac{L^4\omega}{f}\left(qh^y_t+\omega h^x_y\right)=0, \label{eq:HxyEoM}\\
& iq{h^x_y}'+\frac{i\omega}{f}{h^y_t}'-\frac{L^4}{r^4f^2}\left(\omega^2-q^2f\right)h^r_y+\frac{4i\omega L^4A_t'}{r^2f}a_y=0, \label{eq:HryEoM}\\
& \frac{d}{dr}\left[r^2fa_y'+r^2A_t'h^y_t\right]+\frac{i\omega L^4A_t'}{r^2f}h^r_y+\frac{L^4}{r^2f}\left(\omega^2-q^2f\right)a_y=0, \label{eq:ayEoM}
\end{align}
\end{subequations}
where we raise and lower indices of $h_{\mu\nu}$ using the background metric (\ref{eq:RNAdS4solution}) and where a prime denotes a derivative with respect to $r$. The metric fluctuations with mixed indices are more convenient for numerical calculations, as their leading term near the boundary of the spacetime is a constant.

Not all of these equations of motion are linearly independent -- equations (\ref{eq:HytEoM}) and (\ref{eq:HryEoM}) together imply equation (\ref{eq:HxyEoM}) and similarly equations (\ref{eq:HxyEoM}) and (\ref{eq:HryEoM}) together imply equation (\ref{eq:HytEoM}). This is a manifestation of the underlying gauge symmetries of the linear fluctuations which fix the relationship between $h^y_t$ and $h^x_y$ as we will shortly demonstrate. For our purposes, it is convenient to choose $\omega r^2A_t'\times(\ref{eq:HxyEoM})+qr^4f'\times(\ref{eq:ayEoM})+qr^2fA_t'\times(\ref{eq:HytEoM})$, along with equations (\ref{eq:HxyEoM}) and (\ref{eq:HryEoM}), as the three linearly-independent equations of motion. This first combination can be written
\begin{equation}
\label{eq:combinedequation1}
\begin{aligned}
\frac{d}{dr}\left[r^6fA_t'\left({h^y_t}'+\frac{\omega}{q}{h^x_y}'\right)+r^6ff'a_y'\right]&+\frac{L^4r^2A_t'}{f}\left(\omega^2-q^2f\right)\left(h^y_t+\frac{\omega}{q}h^x_y\right)\\
&+\frac{L^4r^2f'}{f}\left(\omega^2-q^2f\right)a_y=0,
\end{aligned}
\end{equation}
where we have used the fact that $r^2A_t'\left(r\right)$ is independent of $r$ and that
\begin{equation}
4L^4A_t'=\frac{d}{dr}\left[\frac{r^2f'}{A_t'}\right],
\end{equation}
for our background. This is a convenient choice because the terms outside the derivative bracket (which are gauge-invariant) are suppressed at sufficiently low $\omega$ and $q$ with respect to the terms inside the derivative bracket (which are also gauge-invariant). Thus at sufficiently low $\omega$ and $q$, this equation (along with equation (\ref{eq:HxyEoM})) can be trivially integrated to produce gauge-invariant first-order differential equations for the fields.

As alluded to above, the linearised fluctuations possess a diffeomorphism gauge symmetry, under which the fields transform as \cite{Wald}
\begin{equation}
\label{eq:DiffeomorphismTransformations}
h_{\mu\nu}\rightarrow h_{\mu\nu}-\nabla_\mu\xi_\nu-\nabla_\nu\xi_\mu,\;\;\;\;\;\;\;\;\;\; a_\mu\rightarrow a_\mu-\xi^\alpha\nabla_\alpha A_\mu-A_\alpha\nabla_\mu\xi^\alpha,
\end{equation}
where the covariant derivatives are taken with respect to the planar RN-AdS$_4$ metric (\ref{eq:RNAdS4solution}). The gauge field fluctuations also transform as $a_\mu\rightarrow a_\mu-\partial_\mu\Lambda$ under the bulk $U(1)$ gauge symmetry, which means that the transverse gauge field $a_y$ is invariant under this transformation, as are all of the metric fluctuations $h_{\mu\nu}$. It will be convenient to write our theory in terms of gauge-invariant combinations of the fluctuations of the fundamental fields $h_{\mu\nu}$ and $a_\mu$. One can construct various different gauge-invariant combinations from the fundamental fields and their derivatives, and the choice we will make is to consider combinations of fields which do not have an $r$ index. It is these fields whose boundary values have a clear interpretation as the sources of the dual field theory operators. With this restriction, the gauge-invariant fields are
\begin{equation}
\label{eq:DefinitionofOurGIFields}
\varphi_1\left(r,\omega,q\right)=h^y_t\left(r,\omega,q\right)+\frac{\omega}{q}h^x_y\left(r,\omega,q\right),\;\;\;\;\;\;\;\;\;\;\;\;\;\;\; \varphi_2\left(r,\omega,q\right)=\frac{L^2}{r_0}a_y\left(r,\omega,q\right),
\end{equation}
or any linear combination of them. When written in terms of these variables, the gravitational theory explicitly encodes the Ward identities of the operators in the dual field theory \cite{Kovtun:2005ev}
\begin{equation}
\label{eq:WardIdentities}
G^R_{T^{xy}T^{ty}}=G^R_{T^{ty}T^{xy}}=\frac{\omega}{q}G^R_{T^{ty}T^{ty}},\;\;\;\;\;\;\;\;G^R_{T^{xy}T^{xy}}=\frac{\omega^2}{q^2}G^R_{T^{ty}T^{ty}},\;\;\;\;\;\;\;\;G^R_{T^{xy}J^{y}}=\frac{\omega}{q}G^R_{T^{ty}J^{y}},
\end{equation}
where these expressions should be understood to hold up to contact terms. This will be seen clearly in the following sections.

To obtain the equations of motion in terms of these variables, one simply solves equation (\ref{eq:HryEoM}) algebraically for $h^r_y$ and then substitutes this solution into the dynamical equations (\ref{eq:combinedequation1}) and (\ref{eq:HxyEoM}). This yields the following coupled equations of motion
\begin{subequations}
\begin{align}
&\frac{d}{dr}\left[r^6f\left(A_t'\varphi_1'+\frac{r_0f'}{L^2}\varphi_2'\right)\right]+\frac{L^4r^2}{f}\left(\omega^2-q^2f\right)\left[A_t'\varphi_1+\frac{r_0f'}{L^2}\varphi_2\right]=0,\label{eq:FinalEoM1}\\
&\frac{d}{dr}\left[\frac{r^2f}{\omega^2-q^2f}\left(r^2\varphi_1'+4r_0L^2A_t'\varphi_2\right)\right]+\frac{L^4}{f}\varphi_1=0, \label{eq:FinalEoM2}
\end{align}
\end{subequations}
which will be the subject of the remainder of this paper.

To compute the Greens functions of the dual field theory, we also need to know the on-shell gravitational action. This can easily be computed in terms of $h_{\mu\nu}$ and $a_\mu$, and after substituting in the solution for $h^r_y$ from equation (\ref{eq:HryEoM}), it becomes
\begin{equation}
\label{eq:OnShellGIAction}
\begin{aligned}
S=\frac{1}{2\kappa_4^2}\int_{r\rightarrow\infty}\frac{d\omega dq}{\left(2\pi\right)^2}\Biggl[&-\frac{r^4fq^2}{2L^4\left(\omega^2-q^2f\right)}\varphi_1\left(r,-\omega,-q\right)\varphi_1'\left(r,\omega,q\right)\\
&-\frac{2r_0^2r^2f}{L^4}\varphi_2\left(r,-\omega,-q\right)\varphi_2'\left(r,\omega,q\right)+\text{non-derivative terms}\Biggr],
\end{aligned}
\end{equation}
where a prime denotes a derivative with respect to $r$. The non-derivative terms in this action produce contact terms in the Greens functions. These non-derivative terms cannot be written purely in terms of the gauge-invariant variables above, which is simply a reflection of the fact that the Ward identities (\ref{eq:WardIdentities}) hold up to contact terms. In the gravitational theory, it is not a signal of broken diffeomorphism invariance but rather that the linearised transformations (\ref{eq:DiffeomorphismTransformations}) should be modified at quadratic order.

\section{Greens functions in the usual hydrodynamic limit $\omega\ll T$}
\label{sec:HydroCalculation}

As a warmup for our low temperature calculations, in this section we will determine the Greens functions in the usual hydrodynamic limit $\omega\ll T$. As expected, we find a diffusion pole with a dispersion relation given by equations (\ref{eq:ExpectedHydroDiffusionResult}) and (\ref{eq:IntroEtaOverS}). Our results are consistent with those of \cite{Ge:2010yc}, although our method is different. Our method can easily be generalised to compute the Greens functions outside of this limit, as we will show in the following sections.

To find the Greens functions, we divide the bulk spacetime into an inner region (a suitably-defined region near the horizon) and an outer region (a suitably-defined region near the boundary). These two regions overlap over a range of $r$ called the matching region. Firstly, we solve the equations of motion in both the inner region and outer region to determine $\varphi_1$ and $\varphi_2$ in the respective regions up to integration constants. After substituting in the solutions $\varphi_1^\text{outer}$ and $\varphi_2^\text{outer}$, the boundary on-shell action (from which the Greens functions are computed) is determined in terms of the integration constants of the outer solutions. These integration constants should be fixed by demanding that the fields are ingoing at the horizon. This is implemented by imposing ingoing boundary conditions on the solutions in the inner region, and then demanding that the solutions $\varphi_{1,2}^\text{inner}$ and $\varphi_{1,2}^\text{outer}$ are consistent in the matching region. Having fixed the integration constants of the outer solutions, we can read off the Green's functions from the on-shell action.

\subsection{The inner region}

We begin by solving the equations of motion (\ref{eq:FinalEoM1}) and (\ref{eq:FinalEoM2}) in the inner region. After diagonalising the two-derivative terms, we expand these equations around the horizon $r=r_0$ (assuming that $\varphi_1\sim\varphi_2$) to obtain
\begin{equation}
\begin{aligned}
\label{eq:HydroInnerExpansion}
&\varphi_1''+\varphi_1'\left[\frac{1}{r-r_0}+\ldots\right]+\varphi_2'\left[\ldots\right]+\varphi_1\left[\frac{L^4\omega^2}{\left(3-Q^2\right)^2r_0^2\left(r-r_0\right)^2}+\ldots\right]+\varphi_2\left[\ldots\right]=0,\\
&\varphi_2''+\varphi_1'\left[\ldots\right]+\varphi_2'\left[\frac{1}{r-r_0}+\ldots\right]+\varphi_2\left[\frac{L^4\omega^2}{\left(3-Q^2\right)^2r_0^2\left(r-r_0\right)^2}+\ldots\right]=0,
\end{aligned}
\end{equation}
where the ellipses denote higher-order terms in this near-horizon expansion. In this region, the fields decouple and their equations of motion can easily be solved to give the usual result for fields near a horizon of non-zero temperature
\begin{equation}
\begin{aligned}
\label{eq:HydroInnerSolutions}
\varphi_{1,2}^\text{inner}&=a^+_{1,2}\left(\frac{r}{r_0}-1\right)^{\frac{i\omega L^2}{r_0\left(3-Q^2\right)}}+a^-_{1,2}\left(\frac{r}{r_0}-1\right)^{-\frac{i\omega L^2}{r_0\left(3-Q^2\right)}}\\
&=a^+_{1,2}\exp\left[\frac{i\omega}{4\pi T}\log\left(\frac{r}{r_0}-1\right)\right]+a^-_{1,2}\exp\left[-\frac{i\omega}{4\pi T}\log\left(\frac{r}{r_0}-1\right)\right],
\end{aligned}
\end{equation}
where $a^\pm_{1,2}$ are integration constants. Ingoing boundary conditions correspond to the choice $a^+_{1,2}=0$.

Note that in the zero temperature limit ($Q\rightarrow\sqrt3$), the expansion (\ref{eq:HydroInnerExpansion})  breaks down, as terms which naively appear to be of a higher order contain factors of the form $\left(3-Q^2\right)^{-1}$. This is simply a consequence of the horizon structure changing in the $T\rightarrow0$ limit as $f(r_0)$ becomes a double zero.

\subsection{The outer region}

We will now solve the equations of motion (\ref{eq:FinalEoM1}) and (\ref{eq:FinalEoM2}) in the outer region. We define this region by
\begin{equation}
\label{eq:HydroOuterLimits1}
\frac{\omega^2L^4}{r^2f^2}\ll1,\;\;\;\;\;\;\;\;\;\;\;\;\frac{q^2L^4}{r^2f}\ll1.
\end{equation}
After taking these limits, the non-derivative terms in the equations of motion drop out and we can trivially integrate them to give
\begin{subequations}
\begin{align}
&r^6f\left(A_t'{\varphi_1^\text{outer}}'+\frac{r_0f'}{L^2}{\varphi_2^\text{outer}}'\right)=\frac{c_1\omega^2r_0}{L^2},\label{eq:HydroOuterEquationsPart1}\\
&\frac{r^2f}{\omega^2-q^2f}\left(r^2{\varphi_1^\text{outer}}'+4r_0L^2A_t'\varphi_2^\text{outer}\right)=\frac{c_2r_0}{L^2r^2A_t'},\label{eq:HydroOuterEquationsPart2}
\end{align}
\end{subequations}
where $c_1$ and $c_2$ are integration constants and we have used the fact that $r^2A_t'$ is a constant for the planar RN-AdS$_4$ background. It is then simple to decouple $\varphi_2^\text{outer}$, which obeys the following linear, first-order differential equation
\begin{equation}
{\varphi_2^\text{outer}}'-\frac{4L^4{A_t'}^2}{r^2f'}\varphi_2^\text{outer}=\frac{\omega^2\left(c_1-c_2\right)+c_2q^2f}{r^6ff'}.
\end{equation}
This equation can be formally solved by means of an integrating factor to give
\begin{equation}
\label{eq:Hydro2OuterSolnBasic}
\varphi_2^\text{outer}\left(r\right)=\exp\left(\int^rd\hat{r}\frac{4L^4{A_t'}^2}{\hat{r}^2f'}\right)\left[b_1+\int^rd\hat{r}\exp\left(-\int^{\hat{r}}d\tilde{r}\frac{4L^4{A_t'}^2}{\tilde{r}^2f'}\right)\frac{\omega^2\left(c_1-c_2\right)+c_2q^2f}{\hat{r}^6ff'}\right].
\end{equation}
where $b_1$ is an integration constant. Substituting this solution for $\varphi_2^\text{outer}$ into equation (\ref{eq:HydroOuterEquationsPart2}), we can easily integrate to obtain $\varphi_1^\text{outer}$
\begin{equation}
\label{eq:Hydro1OuterSolnBasic}
\varphi_1^\text{outer}\left(r\right)=b_2+\int^rd\hat{r}\left(\frac{r_0c_2\left(\omega^2-q^2f\right)}{L^2\hat{r}^6fA_t'}-\frac{4r_0L^2A_t'}{\hat{r}^2}\varphi_2^\text{outer}\left(\hat{r}\right)\right),
\end{equation}
where $b_2$ is an integration constant. Doing the integral
\begin{equation}
\int^rd\hat{r}\frac{4L^4{A_t'}^2}{\hat{r}^2f'}=\log\left[3\left(1+Q^2\right)-4Q^2\frac{r_0}{r}\right],
\end{equation}
we find that
\begin{equation}
\begin{aligned}
\label{eq:Hydro1OuterSoln}
\varphi_1^\text{outer}\left(r\right)&=\varphi_1^{(0)}+\int^rd\hat{r}\frac{r_0c_2\left(\omega^2-q^2f\right)}{L^2\hat{r}^6fA_t'}-\int^rd\hat{r}\frac{4r_0L^2A_t'\varphi_2^{(0)}}{\hat{r}^2}\left(1-\frac{4Q^2}{3\left(1+Q^2\right)}\frac{r_0}{\hat{r}}\right)\\
&-\int^rd\hat{r}\frac{4r_0L^2A_t'}{\hat{r}^2}\left(1-\frac{4Q^2}{3\left(1+Q^2\right)}\frac{r_0}{\hat{r}}\right)\int^{\hat{r}}\frac{d\tilde{r}}{1-\frac{4Q^2}{3\left(1+Q^2\right)}\frac{r_0}{\tilde{r}}}\frac{\omega^2\left(c_1-c_2\right)+c_2q^2f}{\tilde{r}^6ff'},
\end{aligned}
\end{equation}
and
\begin{equation}
\label{eq:Hydro2OuterSoln}
\varphi_2^\text{outer}\left(r\right)=\left(1-\frac{4Q^2}{3\left(1+Q^2\right)}\frac{r_0}{r}\right)\left[\varphi_2^{(0)}+\int^r\frac{d\hat{r}}{1-\frac{4Q^2}{3\left(1+Q^2\right)}\frac{r_0}{\hat{r}}}\frac{\omega^2\left(c_1-c_2\right)+c_2q^2f}{\hat{r}^6ff'}\right],
\end{equation}
where $\varphi^{(0)}_{1,2}$ are the values of $\varphi_{1,2}$ at the boundary of the planar RN-AdS$_4$ spacetime. 

Doing the integrals for the terms proportional to $q^2$, these become
\begin{equation}
\label{eq:Hydro1OuterSolnImproved}
\begin{aligned}
\varphi_1^\text{outer}\left(r\right)=\varphi^{(0)}_1&+\frac{1}{3r^3}\left(1-\frac{Q^2r_0}{\left(1+Q^2\right)r}\right)\left(4Qr_0^3\varphi^{(0)}_2+\frac{c_2q^2}{Qr_0}\right)+\int^rd\hat{r}\frac{r_0\omega^2c_2}{L^2\hat{r}^6fA_t'}\\
&-\int^rd\hat{r}\frac{4r_0L^2A_t'}{\hat{r}^2}\left(1-\frac{4Q^2}{3\left(1+Q^2\right)}\frac{r_0}{\hat{r}}\right)\int^{\hat{r}}d\tilde{r}\frac{\omega^2\left(c_1-c_2\right)}{\tilde{r}^6ff'\left(1-\frac{4Q^2}{3\left(1+Q^2\right)}\frac{r_0}{\tilde{r}}\right)},
\end{aligned}
\end{equation}
and
\begin{equation}
\label{eq:Hydro2OuterSolnImproved}
\begin{aligned}
\varphi_2^\text{outer}\left(r\right)=\left(1-\frac{4Q^2}{3\left(1+Q^2\right)}\frac{r_0}{r}\right)\Biggl[\varphi^{(0)}_2&-\frac{c_2q^2}{3r_0^3r\left(1+Q^2\right)\left(1-\frac{4Q^2}{3\left(1+Q^2\right)}\frac{r_0}{r}\right)}\\
&+\int^rd\hat{r}\frac{\omega^2\left(c_1-c_2\right)}{\hat{r}^6ff'\left(1-\frac{4Q^2}{3\left(1+Q^2\right)}\frac{r_0}{\hat{r}}\right)}\Biggr].
\end{aligned}
\end{equation}
We could not do the integrals exactly for the terms proportional to $\omega^2$ (except in the $T=0$ or $\mu=0$ limits), but this will not be important for determining the Greens function at lowest order in $\omega$ and $q$, as will become clear shortly.

\subsection{Matching}
\label{sec:HydroMatchingSection}

To completely determine the outer solutions and hence the Greens functions, we must fix the integration constants $c_1$ and $c_2$ such that the fields are ingoing at the horizon. As previously mentioned, this can be done by finding a matching region -- a range of $r$ over which both the inner solutions and outer solutions are valid -- and comparing the solutions in that region. As the inner region solutions are valid near the horizon, we expand the outer region solutions near $r=r_0$. For the terms proportional to $\omega^2$, it is sufficient to expand the integrands in this limit and keep the leading terms to give
\begin{equation}
\label{eq:HydroOuterMatchingSolPhi1}
\begin{aligned}
\varphi_1^\text{outer}=\frac{c_2\omega^2}{Qr_0^4\left(3-Q^2\right)}\log\left(\frac{r}{r_0}-1\right)+\Biggl[\varphi^{(0)}_1&+\frac{4Q\varphi^{(0)}_2}{3\left(1+Q^2\right)}+\frac{c_2q^2}{3Qr_0^4\left(1+Q^2\right)}\\
&+O\left(\omega^2c_1,\omega^2c_2\right)\Biggr]+O\left(r-r_0\right),
\end{aligned}
\end{equation}
and
\begin{equation}
\label{eq:HydroOuterMatchingSolPhi2}
\begin{aligned}
\varphi_2^\text{outer}=&\left(1-\frac{4Q^2}{3\left(1+Q^2\right)}\right)\frac{3\left(1+Q^2\right)\omega^2\left(c_1-c_2\right)}{\left(3-Q^2\right)^3r_0^4}\log\left(\frac{r}{r_0}-1\right)\\
&+\left[\varphi_2^{(0)}\left(1-\frac{4Q^2}{3\left(1+Q^2\right)}\right)-\frac{c_2q^2}{3r_0^4\left(1+Q^2\right)}+O\left(\omega^2c_1,\omega^2c_2\right)\right]+O\left(r-r_0\right).
\end{aligned}
\end{equation}
where we have neglected terms of order $c_1\omega^2$ and $c_2\omega^2$ in the coefficients of the $\left(r-r_0\right)^0$ terms. This is because we are searching for a diffusive mode with $\omega\sim q^2$ and thus the $\omega^2c_1$ and $\omega^2c_2$ terms are subleading in an expansion at low $q$. When $r-r_0\ll r_0$, as in these solutions, our constraint on the validity of the outer region solutions (\ref{eq:HydroOuterLimits1}) becomes
\begin{equation}
\frac{\omega L^2}{r_0^2f'(r_0)}\ll\frac{r-r_0}{r_0}\ll1.
\end{equation}
This requires that $\omega\ll T$, which is the hydrodynamic limit commonly used in holographic computations of Greens functions.

The form of the inner solutions (\ref{eq:HydroInnerSolutions}) in the matching region are found by expanding them in the limit
\begin{equation}
\label{eq:HydroInnerLimits2}
\frac{i\omega}{4\pi T}\log\left(\frac{r}{r_0}-1\right)\ll1,
\end{equation}
to give
\begin{equation}
\varphi_{1,2}^\text{inner}\left(r\right)=\left(a^+_{1,2}+a^-_{1,2}\right)+\left(a^+_{1,2}-a^-_{1,2}\right)\frac{i\omega}{4\pi T}\log\left(\frac{r}{r_0}-1\right)+\ldots.
\end{equation}
Demanding ingoing boundary conditions at the horizon ($a^+_{1,2}=0$) fixes the relative coefficient of the constant and logarithmic terms in this expansion
\begin{equation}
\label{eq:HydroInnerMatchingSolutions}
\varphi_{1,2}^\text{inner}\left(r\right)=a^-_{1,2}\left[1-\frac{i\omega}{4\pi T}\log\left(\frac{r}{r_0}-1\right)+\ldots\right].
\end{equation}

To determine $c_1$ and $c_2$ for a solution which is ingoing near the horizon, we equate the coefficients of the constant and logarithmic terms in the outer solutions (\ref{eq:HydroOuterMatchingSolPhi1}), (\ref{eq:HydroOuterMatchingSolPhi2}) with those in the inner solutions (\ref{eq:HydroInnerMatchingSolutions}). The resulting two equations are
\begin{equation}
\begin{aligned}
&\frac{i\omega c_2}{Qr_0^3L^2}=\varphi_1^{(0)}+\frac{4Q\varphi_2^{(0)}}{3\left(1+Q^2\right)}+\frac{c_2q^2}{3Qr_0^4\left(1+Q^2\right)}+O\left(c_1\omega^2,c_2\omega^2\right),\\
&\frac{i\omega\left(c_1-c_2\right)}{L^2r_0^3\left(3-Q^2\right)}=\varphi^{(0)}_2\left(1-\frac{4Q^2}{3\left(1+Q^2\right)}\right)-\frac{c_2q^2}{3r_0^4\left(1+Q^2\right)}+O\left(c_1\omega^2,c_2\omega^2\right).
\end{aligned}
\end{equation}
Neglecting terms of $O\left(\omega^2c_1,\omega^2c_2\right)$, one can solve these to give
\begin{subequations}
\begin{align}
&\begin{aligned}
c_1=\frac{L^2r_0^3}{i\omega\left(i\omega-\frac{L^2q^2}{3r_0\left(1+Q^2\right)}\right)}\Biggl[&Q\varphi^{(0)}_1\left(i\omega-\frac{L^2\left(3-Q^2\right)}{3r_0\left(1+Q^2\right)}q^2\right)\\
&+\frac{\varphi^{(0)}_2}{3\left(1+Q^2\right)}\left(i\omega\left(Q^4-2Q^2+9\right)-\frac{L^2q^2}{r_0}\left(3-Q^2\right)\right)\Biggr],\label{eq:HydroConstants1}
\end{aligned}\\
&c_2=\frac{L^2r_0^3Q\left(\varphi_1^{(0)}+\frac{4Q}{3\left(1+Q^2\right)}\varphi_2^{(0)}\right)}{i\omega-\frac{L^2q^2}{3r_0\left(1+Q^2\right)}}.\label{eq:HydroConstants2}
\end{align}
\end{subequations}
These constants, along with the solutions (\ref{eq:Hydro1OuterSolnImproved}) and (\ref{eq:Hydro2OuterSolnImproved}), determine the behaviour of $\varphi_1$ and $\varphi_2$ near the boundary of the spacetime, assuming that the fields are ingoing at the horizon and that we are in the usual hydrodynamic limit $\omega\ll T$.

\subsection{Green's functions}

To determine the retarded Green's functions, we require the on-shell action for fields that are ingoing near the horizon. Substituting the outer solutions (\ref{eq:Hydro1OuterSolnImproved}) and (\ref{eq:Hydro2OuterSolnImproved}) into the on-shell action (\ref{eq:OnShellGIAction}), we find (where the argument $\left(\omega\right)$ is shorthand for $\left(\omega,q\right)$)
\begin{equation}
\begin{aligned}
S=\frac{1}{2\kappa_4^2}\int_{r\rightarrow\infty}\frac{d\omega dq}{\left(2\pi\right)^2}&\frac{-r_0^2q^2}{2L^2\left(i\omega-\frac{L^2q^2}{3r_0\left(1+Q^2\right)}\right)}\Biggl[\varphi_1^{(0)}\left(-\omega\right)\varphi_1^{(0)}\left(\omega\right)+\frac{16Q^2}{9\left(1+Q^2\right)^2}\varphi_2^{(0)}\left(-\omega\right)\varphi_2^{(0)}\left(\omega\right)\\
&+\frac{4Q}{3\left(1+Q^2\right)}\left(\varphi_1^{(0)}\left(-\omega\right)\varphi_2^{(0)}\left(\omega\right)+\varphi_2^{(0)}\left(-\omega\right)\varphi_1^{(0)}\left(\omega\right)\right)\Biggr],
\end{aligned}
\end{equation}
where we have imposed the values of $c_1$ and $c_2$ for ingoing modes (\ref{eq:HydroConstants1}) and (\ref{eq:HydroConstants2}) and neglected contact terms. Using the relations
\begin{equation}
\label{eq:FieldTheorySourcesInTermsofGIFields}
\varphi_1^{(0)}\left(\omega,q\right)={h^y_t}^{(0)}\left(\omega,q\right)+\frac{\omega}{q}{h^x_y}^{(0)}\left(\omega,q\right),\;\;\;\;\;\;\;\;\;\;\;\;\;\;\;\;\;\;\;\varphi_2^{(0)}\left(\omega,q\right)=\frac{L^2}{r_0}a_y^{(0)}\left(\omega,q\right),
\end{equation}
where the superscript $(0)$ denotes the boundary value of a field, and the usual prescription for calculating Greens functions from holography \cite{Son:2002sd,Kaminski:2009dh}, we find that the retarded Greens functions are (up to contact terms)
\begin{equation}
\label{eq:HydroGreensFunctionsResult}
\begin{aligned}
G^R_{T^{ty}T^{ty}}&=\frac{r_0^2q^2}{2\kappa_4^2L^2\left(i\omega-\frac{L^2q^2}{3r_0\left(1+Q^2\right)}\right)},\;\;\;\;\;\;\;\;\;\;\;G^R_{J^yJ^y}=\frac{8L^2q^2Q^2}{9\kappa_4^2\left(1+Q^2\right)^2\left(i\omega-\frac{L^2q^2}{3r_0\left(1+Q^2\right)}\right)},\\
G^R_{T^{ty}J^y}&=G^R_{J^{y}T^{ty}}=\frac{2r_0q^2Q}{3\kappa_4^2\left(1+Q^2\right)\left(i\omega-\frac{L^2q^2}{3r_0\left(1+Q^2\right)}\right)},
\end{aligned}
\end{equation}
with the Greens functions of the other operators obeying the Ward identities (\ref{eq:WardIdentities}) due to our gauge-invariant bulk formulation. These results are valid only at lowest order in $\omega$ and $q$ (assuming that $\omega\sim q^2$), and are valid only up to contact terms. They agree with the results of \cite{Ge:2010yc} up to such corrections.\footnote{Although the correlators involving $J^y$ appear superficially different from those in \cite{Ge:2010yc}, these differences are just due to contact terms. Both our calculation and the calculation in \cite{Ge:2010yc} neglect certain contact terms, but not the same set of contact terms (since they are done by different methods).} They accurately capture the behaviour of the spectral function in the neighbourhood of the diffusion peak, as we will show numerically in section \ref{sec:NumericalResults}.

The Greens functions share a common diffusive pole with dispersion relation
\begin{equation}
\label{eq:HydroDerivedDispersionRelationResult}
\omega=-i\frac{L^2}{3r_0\left(1+Q^2\right)}q^2+\ldots=-i\frac{s}{4\pi\left(\epsilon+P\right)}q^2+\ldots,
\end{equation}
where the ellipses denote higher order terms in $q$, as one expects from the hydrodynamic equations (\ref{eq:ExpectedHydroDiffusionResult}) and (\ref{eq:IntroEtaOverS}). This value of the viscosity can be checked by applying the Kubo formula (\ref{eq:KuboFormulaForViscosity}) to our results for the Greens functions.

\section{Zero temperature Greens functions}
\label{sec:ZeroTCalculation}

As we emphasised, the above calculation of the Greens functions is not valid when $T=0$. This is because the horizon structure is qualitatively different in this limit -- $f(r)$ has a double zero when $T=0$, rather than the single zero it has when $T\ne0$. However, at $T=0$ one can use a similar procedure to that just outlined provided that the inner region is suitably defined. For the extremal background there is a natural candidate for such an inner region -- the AdS$_2\times \mathbb{R}^2$ near-horizon geometry.

\subsection{The inner AdS$_2$ region}

Following similar calculations in \cite{Faulkner:2009wj,Edalati:2009bi,Edalati:2010hk,Edalati:2010pn,Faulkner:2012gt}, the near-horizon AdS$_2\times \mathbb{R}^2$ geometry is most easily seen by changing co-ordinates to
\begin{equation}
\label{eq:DefinitionOfZeta}
\zeta=\frac{r-r_0}{\omega L^2},
\end{equation}
and then expanding linear combinations of the equations of motion (\ref{eq:FinalEoM1}) and (\ref{eq:FinalEoM2}) as power series at small $\omega$ and $q$  (assuming that $\varphi_1\sim\varphi_2$) to give
\begin{equation}
\begin{aligned}
\label{eq:InnerRegionEquationsofMotionZeroT}
&\varphi_1''\left(\zeta\right)+\varphi_1'\left(\zeta\right)\left[\frac{2}{\zeta}+\ldots\right]+\varphi_2'\left(\zeta\right)\left[\ldots\right]+\varphi_1\left(\zeta\right)\left[\frac{1}{36\zeta^4}+\ldots\right]+\varphi_2\left(\zeta\right)\left[\ldots\right]=0,\\
&\varphi_2''\left(\zeta\right)+\varphi_1'\left(\zeta\right)\left[\ldots\right]+\varphi_2'\left(\zeta\right)\left[\frac{2}{\zeta}+\ldots\right]+\varphi_2\left(\zeta\right)\left[\frac{1}{36\zeta^4}-\frac{2}{\zeta^2}+\ldots\right]=0,
\end{aligned}
\end{equation}
where the ellipses denote terms with positive powers of $\omega,q$. In these limits, the equations of motion decouple and after redefining the radial variable $\zeta\rightarrow\hat{\zeta}/\left(\omega L^2\right)$ to remove the frequency dependence, the equations for $\varphi_1^\text{inner}(\hat{\zeta})$ and $\varphi_2^\text{inner}(\hat{\zeta})$ are simply the equations of motion for scalar fields in AdS$_2$ with $\left(mL_2\right)^2=0,2$ respectively. The radius of curvature of this effective AdS$_2$ spacetime is related to the radius of curvature of the RN-AdS$_4$ spacetime via $L_2=L/\sqrt{6}$. 

It is simple to solve these inner region equations to give
\begin{equation}
\label{eq:ZeroTInnerRegionSolutions}
\begin{aligned}
\varphi_1^\text{inner}\left(\zeta\right)&=a_1^+\exp\left(\frac{i}{6\zeta}\right)+a_1^-\exp\left(-\frac{i}{6\zeta}\right),\\
\varphi_2^\text{inner}\left(\zeta\right)&=a_2^+\left(\zeta-\frac{i}{6}\right)\exp\left(\frac{i}{6\zeta}\right)+a_2^-\left(\zeta+\frac{i}{6}\right)\exp\left(-\frac{i}{6\zeta}\right),
\end{aligned}
\end{equation}
where $a^\pm_{1,2}$ are integration constants. The equations (\ref{eq:InnerRegionEquationsofMotionZeroT}) are valid in the limit of small frequency and momentum (with respect to $r_0/L^2$) with $\zeta$, defined in (\ref{eq:DefinitionOfZeta}), fixed. These requirements imply that $r-r_0\ll r_0$ i.e. that these are near-horizon solutions. Ingoing boundary conditions at the horizon ($\zeta\rightarrow0$) correspond to the choice $a_{1,2}^-=0$. At $T=0$, $r_0/L^2$ is essentially the chemical potential $\mu$ of the field theory and so the following results are valid only for $\omega,q\ll\mu$.

\subsection{The outer region}

As in the $T\ne0$ case, we define the outer region by (\ref{eq:HydroOuterLimits1}). Just as at $T\ne0$, the equations of motion in the outer region can then be solved to give the integral solutions (\ref{eq:Hydro1OuterSolnImproved}) and (\ref{eq:Hydro2OuterSolnImproved}), with $Q=\sqrt{3}$ for the $T=0$ case. In the limit $T=0$, the fact that $f(r)$ has a double pole means that these integrals can be done exactly but the results are lengthy and will not be presented here. To compute the Greens functions, we only require to know the behaviour of the outer solutions in the matching region and near the boundary of the spacetime, and this will be presented in the following subsections.

\subsection{Matching}

To fix the integration constants $c_1$ and $c_2$ in the outer region solutions, we need to know the inner and outer solutions in the matching region where the inner and outer regions overlap. The solutions overlap if we expand the inner solutions (\ref{eq:ZeroTInnerRegionSolutions}) in the limit $\zeta\rightarrow\infty$ (`far from the horizon') to obtain
\begin{equation}
\label{eq:ZeroTInnerMatchingExpansions}
\begin{aligned}
\varphi_1^\text{inner}&=\left(a^+_1+a^-_1\right)\left[1+O\left(\zeta^{-2}\right)\right]+\frac{i}{6\zeta}\left(a^+_1-a^-_1\right)\left[1+O\left(\zeta^{-2}\right)\right],\\
\varphi_2^\text{inner}&=\zeta\left(a^+_2+a^-_2\right)\left[1+O\left(\zeta^{-2}\right)\right]+\frac{i}{648\zeta^2}\left(a^+_2-a^-_2\right)\left[1+O\left(\zeta^{-2}\right)\right],
\end{aligned}
\end{equation}
and the outer solutions (\ref{eq:Hydro1OuterSolnImproved}) and (\ref{eq:Hydro2OuterSolnImproved}) in the limit $r\rightarrow r_0$ (`near the horizon')
\begin{equation}
\label{eq:ZeroTOuterSolutionsMatchingSeries}
\begin{aligned}
\varphi_1^\text{outer}&=-\frac{\left(c_1+2c_2\right)\omega^2}{18\sqrt{3}r_0^3\left(r-r_0\right)}+\ldots+\left[\varphi_1^{(0)}+\frac{1}{\sqrt{3}}\varphi_2^{(0)}+\frac{\sqrt{3}c_2q^2}{36r_0^4}+O\left(\omega^2c_1,\omega^2c_2\right)\right]+\ldots,\\
\varphi_2^\text{outer}&=-\frac{\omega^2\left(c_1-c_2\right)}{216r_0^2\left(r-r_0\right)^2}+\ldots+\left[\frac{1}{r_0}\varphi_2^{(0)}+\frac{c_2q^2}{12r_0^5}+O\left(\omega^2c_1,\omega^2c_2\right)\right]\left(r-r_0\right)+\ldots,
\end{aligned}
\end{equation}
where we have only explicitly written the terms required for the matching. As was the case for $\omega\ll T$, we are searching for a diffusion mode with $\omega\sim q^2$ and thus we can discard the $O\left(\omega^2c_1,\omega^2c_2\right)$ terms as they are subleading at low $q$.

After changing co-ordinates back from $\zeta$ to $r$ in (\ref{eq:ZeroTInnerMatchingExpansions}), imposing ingoing boundary conditions fixes the ratio of the coefficients of the $\left(r-r_0\right)^{-1}$ and the $\left(r-r_0\right)^{0}$ terms of $\varphi_1^\text{inner}$ in the matching region, and the ratio of the coefficients of the $\left(r-r_0\right)^{-2}$ and the $\left(r-r_0\right)^{1}$ terms of $\varphi_2^\text{inner}$ in the matching region as follows 
\begin{equation}
\label{eq:ZeroTInnerSolutionsMatchingSeries}
\begin{aligned}
\varphi_1^\text{inner}&=a_1^+\left[1+\mathcal{G}^\text{IR}_{1}\left(\omega\right)\frac{1}{r-r_0}+\ldots\right],\\
\varphi_2^\text{inner}&=a_2^+\left[\left(r-r_0\right)+\ldots+\mathcal{G}^\text{IR}_{2}\left(\omega\right)\frac{1}{\left(r-r_0\right)^2}+\ldots\right],
\end{aligned}
\end{equation}
where we have absorbed a factor of $\omega L^2$ into $a_2^+$ and where
\begin{equation}
\label{eq:IRGreensFunctionDefinitions}
\mathcal{G}^\text{IR}_{1}\left(\omega\right)=\frac{i\omega L^2}{6},\;\;\;\;\;\;\;\;\;\;\;\;\;\;\;\;\;\mathcal{G}^\text{IR}_{2}\left(\omega\right)=\frac{i\omega^3 L^6}{648},
\end{equation}
are (proportional to) the retarded Greens functions of scalar operators with conformal dimensions $\Delta=1,2$ in the CFT$_1$ dual to the near-horizon AdS$_2$ geometry respectively. The relative coefficients of the terms in each series are determined by these CFT$_1$ Greens functions simply because $\varphi_{1}^\text{inner}$ and $\varphi_{2}^\text{inner}$ behave like scalar fields in AdS$_2$ with masses $\left(mL_2\right)^2=0,2$ respectively. It is instructive for now to work with $\mathcal{G}^\text{IR}_{1}\left(\omega\right)$ and $\mathcal{G}^\text{IR}_{2}\left(\omega\right)$, rather than using their explicit values (\ref{eq:IRGreensFunctionDefinitions}).

To determine $c_1$ and $c_2$ for solutions which are ingoing at the horizon, we fix the coefficients of the power series in the outer solutions (\ref{eq:ZeroTOuterSolutionsMatchingSeries}) so that they agree with those of the inner solutions in the matching region (\ref{eq:ZeroTInnerSolutionsMatchingSeries}). Neglecting the $O\left(\omega^2c_1,\omega^2c_2\right)$ terms, which are subleading at low $q$ as previously explained, this fixes
\begin{equation}
\begin{aligned}
c_1&=\frac{4\sqrt{3}r_0^3\left[24\sqrt{3}r_0\omega^2\varphi_2^{(0)}\mathcal{G}^\text{IR}_{2}+\mathcal{G}^\text{IR}_{1}\left(-54\mathcal{G}^\text{IR}_{2}q^2\varphi_1^{(0)}+r_0^3\omega^2\left[\sqrt{3}\varphi_2^{(0)}+3\varphi_1^{(0)}\right]\right)\right]}{\omega^2\left[12\mathcal{G}^\text{IR}_{2}q^2-r_0^2\left(\mathcal{G}^\text{IR}_{1}q^2+2r_0\omega^2\right)\right]},\\
c_2&=\frac{12r_0^4\left[-12\mathcal{G}^\text{IR}_{2}\varphi_2^{(0)}+r_0^2\mathcal{G}^\text{IR}_{1}\left(\sqrt{3}\varphi_1^{(0)}+\varphi_2^{(0)}\right)\right]}{12\mathcal{G}^\text{IR}_{2}q^2-r_0^2\left(\mathcal{G}^\text{IR}_{1}q^2+2r_0\omega^2\right)}.
\end{aligned}
\end{equation}
Due to our previous approximations, these results are only valid to lowest order in $\omega\sim q^2$. Recall that $\mathcal{G}^\text{IR}_{2}\left(\omega\right)$ is suppressed with respect to $\mathcal{G}^\text{IR}_{1}\left(\omega\right)$ by a factor of $\omega^2$ and therefore if we keep only the lowest order terms in these expressions (as we should for consistency), all dependence upon $\mathcal{G}^\text{IR}_{2}\left(\omega\right)$ drops out to give
\begin{equation}
\label{eq:TZeroSolutionsForOuterConstants}
c_1=c_2=-\frac{6r_0^3\mathcal{G}^\text{IR}_1\left(\omega\right)\left(\sqrt{3}\varphi_1^{(0)}+\varphi_2^{(0)}\right)}{\omega^2+\frac{q^2}{2r_0}\mathcal{G}^\text{IR}_1\left(\omega\right)}.
\end{equation}
With these values for $c_1$ and $c_2$, the outer solutions correspond to fields which are ingoing at the horizon.

\subsection{Greens functions}

To determine the retarded Greens functions at $T=0$, we substitute the outer solutions (\ref{eq:Hydro1OuterSolnImproved}) and (\ref{eq:Hydro2OuterSolnImproved}) (with $Q=\sqrt{3}$ and $c_1$ and $c_2$ given by (\ref{eq:TZeroSolutionsForOuterConstants})) into the on-shell boundary action (\ref{eq:OnShellGIAction}). This yields (where the argument $\left(\omega\right)$ is shorthand for $\left(\omega,q\right)$)
\begin{equation}
\begin{aligned}
S=\frac{1}{2\kappa_4^2}\int_{r\rightarrow\infty}\frac{d\omega dq}{\left(2\pi\right)^2}\frac{3r_0^2q^2\mathcal{G}^\text{IR}_1\left(\omega\right)}{L^4\left(\omega^2+\frac{q^2}{2r_0}\mathcal{G}^\text{IR}_1\left(\omega\right)\right)}\Biggl[&\varphi_1^{(0)}\left(-\omega\right)\varphi_1^{(0)}\left(\omega\right)+\frac{1}{3}\varphi_2^{(0)}\left(-\omega\right)\varphi_2^{(0)}\left(\omega\right)\\
&+\frac{1}{\sqrt{3}}\left(\varphi_1^{(0)}\left(-\omega\right)\varphi_2^{(0)}\left(\omega\right)+\varphi_2^{(0)}\left(-\omega\right)\varphi_1^{(0)}\left(\omega\right)\right)\Biggr],
\end{aligned}
\end{equation}
up to contact terms. By using the relations (\ref{eq:FieldTheorySourcesInTermsofGIFields}) and then applying the usual holographic prescriptions \cite{Son:2002sd,Kaminski:2009dh}, we find that the retarded Greens functions are
\begin{equation}
\label{eq:ZeroTGreensFunctionsResults}
\begin{aligned}
G^R_{T^{ty}T^{ty}}&=-\frac{3r_0^2q^2\mathcal{G}^\text{IR}_1\left(\omega\right)}{L^4\kappa_4^2\left(\omega^2+\frac{q^2}{2r_0}\mathcal{G}^\text{IR}_1\left(\omega\right)\right)}=\frac{r_0^2q^2}{2L^2\kappa_4^2\left(i\omega-\frac{q^2L^2}{12r_0}\right)},\\
G^R_{J^{y}J^{y}}&=-\frac{q^2\mathcal{G}^\text{IR}_1\left(\omega\right)}{\kappa_4^2\left(\omega^2+\frac{q^2}{2r_0}\mathcal{G}^\text{IR}_1\left(\omega\right)\right)}=\frac{L^2q^2}{6\kappa_4^2\left(i\omega-\frac{q^2L^2}{12r_0}\right)},\\
G^R_{T^{ty}J^{y}}&=G^R_{J^{y}T^{ty}}=-\frac{\sqrt{3}r_0q^2\mathcal{G}^\text{IR}_1\left(\omega\right)}{L^2\kappa_4^2\left(\omega^2+\frac{q^2}{2r_0}\mathcal{G}^\text{IR}_1\left(\omega\right)\right)}=\frac{\sqrt{3}r_0q^2}{6\kappa_4^2\left(i\omega-\frac{q^2L^2}{12r_0}\right)},
\end{aligned}
\end{equation}
and the Ward identities (\ref{eq:WardIdentities}) are again trivially satisfied due to our gauge-invariant formulation. These results are valid to leading order at low $\omega$ and $q$ (with $\omega\sim q^2$), and are valid only up to contact terms.

The expressions (\ref{eq:ZeroTGreensFunctionsResults}) are precisely those obtained by taking the naive $T=0$ limit of the hydrodynamic results (\ref{eq:HydroGreensFunctionsResult})! In particular, they support a diffusion pole with dispersion relation given by (where the ellipses denote higher order terms in a small $\omega,q$ expansion)
\begin{equation}
\label{eq:ZeroTResultforDispersionRelation}
\omega^2+\frac{q^2}{2r_0}\mathcal{G}^\text{IR}_1\left(\omega\right)+\ldots=0\;\;\;\;\;\;\;\;\;\longrightarrow\;\;\;\;\;\;\;\;\;\omega=-i\frac{L^2}{12r_0}q^2+\ldots,
\end{equation}
which is simply the $T=0$ limit of the hydrodynamic relations (\ref{eq:ExpectedHydroDiffusionResult}) and (\ref{eq:IntroEtaOverS}). As mentioned in the introduction, the main features of this dispersion relation are consistent with the principles apparent from previous studies of the low-energy physics of holographic theories \cite{Iqbal:2008by,Faulkner:2009wj,Faulkner:2010tq,Nickel:2010pr}. That is, the low-energy physics will be dominated by any gapless modes present in the theory in addition to the near-horizon geometry. The existence of such gapless modes is not apparent from the near-horizon geometry only but requires knowledge of the full bulk spacetime, although the decay rate of these modes is controlled by the near-horizon geometry. This is seen in the dispersion relation (\ref{eq:ZeroTResultforDispersionRelation}) above -- the existence of the diffusive mode required knowledge of the outer region of the geometry, but the power of momentum in the dispersion relation is controlled by the conformal dimension of the operator dual to the field $\varphi_1$ in the CFT$_1$ dual to the near-horizon AdS$_2$ geometry. Unlike in the case of probe Dirac fermions studied in \cite{Faulkner:2009wj}, this dimension is fixed.

It is instructive to compare our results to those of \cite{Edalati:2009bi,Edalati:2010hk}, which analytically computed the leading behaviour of the retarded Greens functions in the limit $\omega\rightarrow0$. Their calculation involves a matching procedure similar to that used here, but is formulated in terms of the gauge-invariant Kodama-Ishibashi fields -- for which the equations of motion decouple \cite{Kodama:2003kk} -- rather than the fields $\varphi_1$ and $\varphi_2$ used here. With these decoupled fields, one can determine the solutions in the near-horizon region more accurately than we have done here -- in particular, the conformal dimensions of the CFT$_1$ operators dual to these fields in the near-horizon region are given by
\begin{equation}
\delta_\pm=\frac{1}{2}+\frac{1}{2}\sqrt{5+2\frac{q^2}{\mu^2}\pm4\sqrt{1+\frac{q^2}{\mu^2}}}.
\end{equation}
These dimensions include $q/\mu$-dependent corrections to the values $1$ and $2$ which we used in our calculation. The authors of \cite{Edalati:2010hk} can therefore compute the $\omega\rightarrow0$ scaling behaviour of the imaginary parts of the retarded Greens functions to find, for example,
\begin{equation}
\label{eq:JottarEtAlResultBranchCutTerms}
\text{Im}G^R_{T^{ty}T^{ty}}\propto\left(\frac{\omega}{\mu}\right)^{2\delta_--1}=\frac{\omega}{\mu}+\frac{\omega q^4}{4\mu^5}\log\frac{\omega}{\mu}+\ldots,
\end{equation}
where we have expanded in the limit $q\ll\mu$, compared to our result $\text{Im}G^R_{T^{ty}T^{ty}}\propto\omega/\mu$. 

Using the Kodama-Ishibashi variables also allows allows one to consistently include higher-order frequency corrections to the Greens functions which are analogous to the $\mathcal{G}^\text{IR}_2\left(\omega\right)$ terms we neglected in our derivation. It is this contribution which gives a low-frequency dissipative conductivity (ignoring the delta function peak) of $\sigma\left(\omega\rightarrow0\right)\sim\omega^2$ \cite{Edalati:2009bi}. Our results (\ref{eq:ZeroTGreensFunctionsResults}) yield $\sigma\left(\omega\right)=0$ since $G^R_{J^yJ^y}\left(\omega,q=0\right)$ vanishes due to our neglection of these higher-order terms. Despite these drawbacks of our approach as compared to using the Kodama-Ishibashi variables, the advantage is that we can solve the equations of motion in the outer region more precisely than has been previously done. As we have seen, it is this detailed knowledge of the fields far from the horizon which allows us to determine the existence of the diffusion mode. Of course, our results (\ref{eq:ZeroTGreensFunctionsResults}) give a `zero temperature viscosity' equivalent to that found in \cite{Edalati:2009bi}
\begin{equation}
\eta\equiv-\lim_{\omega\rightarrow0}\frac{1}{\omega}\text{Im}G^R_{T^{xy}T^{xy}}\left(\omega,q=0\right)=\frac{3r_0^2}{L^4\kappa_4^2}\lim_{\omega\rightarrow0}\frac{1}{\omega}\text{Im}\mathcal{G}^\text{IR}_1\left(\omega\right)=\frac{s}{4\pi}.
\end{equation}

\section{Greens functions at low, non-zero temperatures $T\lesssim\omega\ll\mu$}
\label{sec:NonHydroNonZeroTCalculation}

In this section, we will calculate the Greens functions (at leading order in small $\omega$ and $q$) at small (with respect to $\mu$), non-zero temperatures which are outside of the usual hydrodynamic limit. In other words, in the range $0<T\lesssim\omega\ll\mu$. Heuristically, one would expect the calculation to proceed in a similar vein to the $T=0$ calculation but to now include small $T/\mu$ and $T/\omega$ corrections. We will see that this is the case -- for example the near-horizon AdS$_2$ geometry is replaced by a near-horizon Schwarzschild-AdS$_2$ geometry -- and that the leading order Greens functions derived in this way are, in fact, equivalent to those obtained from the hydrodynamic calculation (\ref{eq:HydroGreensFunctionsResult}).

\subsection{The inner Schwarzschild-AdS$_2$ region}

To define the inner region when $T=0$, we wrote the equations in terms of the new radial co-ordinate $\zeta$ (defined in equation (\ref{eq:DefinitionOfZeta})) and then kept only the lowest order terms in $\omega$ and $q$ in these equations. This corresponded to a near-horizon limit at small $\omega$ and $q$. To determine the leading $T$-dependent corrections to this result, we must keep $T\sim\omega$ in this limit also. To this end, we can replace $Q$ (which determines $T/\mu$ via equation (\ref{eq:QasfunctionofTandMu})) with 
\begin{equation}
\zeta_0\equiv\frac{r_0\left(3-Q^2\right)}{\omega L^2}=\frac{4\pi T}{\omega},
\end{equation}
and keep this fixed in the low $\omega$ and $q$ limit. Expanding linear combinations of the equations of motion (\ref{eq:FinalEoM1}) and (\ref{eq:FinalEoM2}) at low $\omega$ and $q$, we find
\begin{equation}
\begin{aligned}
&\varphi_1''\left(\zeta\right)+\varphi_1'\left(\zeta\right)\left[\frac{12\zeta+\zeta_0}{\zeta\left(6\zeta+\zeta_0\right)}+\ldots\right]+\varphi_2'\left(\zeta\right)\left[\ldots\right]+\varphi_2\left(\zeta\right)\left[\ldots\right]+\varphi_1\left(\zeta\right)\left[\frac{1}{\zeta^2\left(6\zeta+\zeta_0\right)^2}+\ldots\right]=0,\\
&\varphi_2''\left(\zeta\right)+\varphi_1'\left(\zeta\right)\left[\ldots\right]+\varphi_2'\left(\zeta\right)\left[\frac{12\zeta+\zeta_0}{\zeta\left(6\zeta+\zeta_0\right)}+\ldots\right]+\varphi_2\left(\zeta\right)\left[\frac{1-72\zeta^2-12\zeta\zeta_0}{\zeta^2\left(6\zeta+\zeta_0\right)^2}+\ldots\right]=0,
\end{aligned}
\end{equation}
where the ellipses denote terms with positive powers of $\omega,q$. As one might expect, in these limits the equations of motion decouple and are simply the equations of motion of scalars of masses $\left(mL_2\right)^2=0,2$ in a Schwarzschild-AdS$_2$ geometry respectively. This is most easily seen by changing co-ordinates to $\zeta=\left(\rho-\rho_0\right)/\omega L^2$ with $\rho_0=\omega L^2\zeta_0/12$, which yields the relevant Schwarzschild-AdS$_2$ scalar equations of motion in a co-ordinate system in which the emblackening factor has the form $f\left(\rho\right)=1-\rho_0^2/\rho^2$.

Solving these equations of motion, we find
\begin{equation}
\begin{aligned}
\label{eq:NonHydroNonZeroTInnerSolutions}
\varphi_1^\text{inner}\left(\zeta\right)&=a_1^{+}\exp\left[\frac{i}{\zeta_0}\log\left(1+\frac{\zeta_0}{6\zeta}\right)\right]+a_1^-\exp\left[-\frac{i}{\zeta_0}\log\left(1+\frac{\zeta_0}{6\zeta}\right)\right],\\
\varphi_2^\text{inner}\left(\zeta\right)&=a_2^+\left(\zeta+\frac{\zeta_0}{12}-\frac{i}{6}\right)\exp\left[\frac{i}{\zeta_0}\log\left(1+\frac{\zeta_0}{6\zeta}\right)\right]+a_2^-\left(\zeta+\frac{\zeta_0}{12}+\frac{i}{6}\right)\exp\left[-\frac{i}{\zeta_0}\log\left(1+\frac{\zeta_0}{6\zeta}\right)\right].\\
\end{aligned}
\end{equation}
By an analogous argument to that at $T=0$, these solutions are valid in the near-horizon region for small $T\sim\omega\sim q\ll r_0/L^2\sim\mu$. Note that in the zero temperature limit $\zeta_0\rightarrow0$, one can expand the logarithms and reproduce the $T=0$ results (\ref{eq:ZeroTInnerRegionSolutions}) found previously. To impose ingoing boundary conditions, we can expand these inner solutions near the Schwarzschild-AdS$_2$ horizon $\zeta\rightarrow0$ to give
\begin{equation}
\varphi_{1,2}^\text{inner}\sim a_{1,2}^+\left(\frac{r}{r_0}-1\right)^{-\frac{i\omega L^2}{r_0\left(3-Q^2\right)}}+a_{1,2}^-\left(\frac{r}{r_0}-1\right)^{\frac{i\omega L^2}{r_0\left(3-Q^2\right)}},
\end{equation}
where the $\sim$ indicates that we have absorbed various $r$-independent factors into the coefficients $a_{1,2}^\pm$. This is the expected result for fields at a horizon at non-zero temperature (recall for example equation (\ref{eq:HydroInnerSolutions})). Ingoing boundary conditions correspond to the choice $a_{1,2}^-=0$.

\subsection{The outer region}

As in the previous cases, we define the outer region by the inequalities (\ref{eq:HydroOuterLimits1}). We have already derived the solutions for $\varphi_{1,2}^\text{outer}$ in this region -- they are given by equations (\ref{eq:Hydro1OuterSolnImproved}) and (\ref{eq:Hydro2OuterSolnImproved}). 

\subsection{Matching}

Finding an overlapping region to match the inner and outer solutions is more involved in this case than in the previous cases and so we proceed carefully. Firstly, we expand the inner solutions (\ref{eq:NonHydroNonZeroTInnerSolutions}) near the boundary of the Schwarzschild-AdS$_2$ spacetime ($\zeta\rightarrow\infty$) to give
\begin{equation}
\begin{aligned}
\varphi_1^\text{inner}\left(\zeta\right)&=\left(a_1^++a_1^-\right)+\frac{i}{6\zeta}\left(a_1^+-a_1^-\right)\left[1+\ldots\right]+\ldots,\\
\varphi_2^\text{inner}\left(\zeta\right)&=\zeta\left(a_2^++a_2^-\right)+\ldots+\frac{i}{2592\zeta^2}\left[a_2^+\left(4+3i\zeta_0+\zeta_0^2\right)-a_2^-\left(4-3i\zeta_0+\zeta_0^2\right)\right]+\ldots,
\end{aligned}
\end{equation}
where we have explicitly written only the leading terms whose ratio of coefficients is fixed by imposing ingoing boundary conditions at the horizon. After imposing ingoing boundary conditions at the horizon ($a_{1,2}^-=0$), absorbing constants into $a_{1,2}^+$ and changing back to the usual radial co-ordinate, we find that near the Schwarzschild-AdS$_2$ boundary,
\begin{equation}
\label{eq:NonZeroTNonHydroInnerMatchingSolutions}
\begin{aligned}
\varphi_1^\text{inner}&=a_1^+\left[1+\mathcal{G}^\text{IR}_{1,T}\left(\omega\right)\frac{1}{r-r_0}+\ldots\right],\\
\varphi_2^\text{inner}&=a_2^+\left[\left(r-r_0\right)+\ldots+\mathcal{G}^\text{IR}_{2,T}\left(\omega\right)\frac{1}{\left(r-r_0\right)^2}+\ldots\right],
\end{aligned}
\end{equation}
where
\begin{equation}
\mathcal{G}^\text{IR}_{1,T}\left(\omega\right)=\frac{i\omega L^2}{6},\;\;\;\;\;\;\;\;\;\;\;\;\;\mathcal{G}^\text{IR}_{2,T}\left(\omega\right)=\frac{i\omega^3L^6}{648}\left(1+\frac{3i\pi T}{\omega}+\frac{4\pi^2T^2}{\omega^2}\right),
\end{equation}
are (proportional to) the retarded Greens functions of scalar operators with conformal dimensions $\Delta=1,2$ in the thermal CFT$_1$ dual to the near-horizon Schwarzschild-AdS$_2$ geometry respectively. As in the $T=0$ case, the relative coefficients of the terms in each series are determined by these Greens functions simply because $\varphi_{1}^\text{inner}$ and $\varphi_{2}^\text{inner}$ behave like scalar fields in Schwarzschild-AdS$_2$ with masses $\left(mL_2\right)^2=0,2$ respectively. 

Note that for a dimension $1$ operator the Greens function is unchanged when one turns on the temperature $T$, whereas the Greens function of the dimension $2$ operator receives non-zero $T/\omega$ corrections. These $T/\omega$ corrections have the potential to introduce $q$-independent (but $T$-dependent) terms into the dispersion relation $\omega\left(q\right)$ at low temperatures as we will shortly show.\footnote{For a toy model of probe D-brane systems similar to those of \cite{Karch:2008fa,Davison:2011ek}, it is similar $T/\omega$ corrections to the inner region solution which result in the Fermi liquid-like zero sound dispersion relation $-\text{Im}\left(\omega\right)\sim k^2+T^2$ \cite{DBraneWorkInProgress}.} However, as in the $T=0$ case, the $\mathcal{G}^\text{IR}_{2,T}$ term is suppressed by a factor of $\omega^2$ with respect to the $\mathcal{G}^\text{IR}_{1,T}$ contribution (recall that we assume $\omega\gtrsim T$ in this section) and thus makes only a subleading contribution to the dispersion relation. It is for this reason that the dispersion relation does not acquire any significant $q$-independent (but $T$-dependent) contributions.

To show this, we must match these inner solutions to the outer solutions over a range of $r$ where both solutions are valid. Simply expanding the outer solutions around $r=r_0$ yields the results (\ref{eq:HydroOuterMatchingSolPhi1}) and (\ref{eq:HydroOuterMatchingSolPhi2}) which do not overlap with the inner solutions (\ref{eq:NonZeroTNonHydroInnerMatchingSolutions}). This is because, as discussed in section \ref{sec:HydroMatchingSection}, these solutions are valid in the usual hydrodynamic limit $\omega\ll T$ whereas the inner solutions (\ref{eq:NonZeroTNonHydroInnerMatchingSolutions}) are valid for $\omega\gtrsim T$. The correct procedure is to expand the outer solutions in the limits $r-r_0\ll r_0$ \textit{and} $3-Q^2\ll1$, where this second limit is equivalent to $T/\mu\ll1$, which is also the range of validity of the inner solutions (\ref{eq:NonZeroTNonHydroInnerMatchingSolutions}). In these limits,
\begin{equation}
\begin{aligned}
f(r)\longrightarrow\frac{1}{r^4}\left(r-r_0\right)^2\left(r^2+2rr_0+3r_0^2\right)=f_0\left(r\right),
\end{aligned}
\end{equation}
where the subscript $0$ indicates that $f(r)$ takes on its $T=0$ form, and
\begin{equation}
1-\frac{4Q^2r_0}{3\left(1+Q^2\right)r}\rightarrow\frac{4Q^2}{3\left(1+Q^2\right)}\left(1-\frac{r_0}{r}\right).
\end{equation}
Note that in these limits, the outer region inequality (\ref{eq:HydroOuterLimits1}) no longer implies that $\omega\ll T$.

With these substitutions, the outer solutions (\ref{eq:Hydro1OuterSolnImproved}) and (\ref{eq:Hydro2OuterSolnImproved}) can be expanded near the horizon $r=r_0$, where they overlap with the inner matching solutions (\ref{eq:NonZeroTNonHydroInnerMatchingSolutions}) just derived. In this matching region, the outer solutions take the form
\begin{equation}
\begin{aligned}
\varphi_1^\text{outer}=&-\frac{\omega^2\left[c_1Q^2+c_2\left(9-Q^2\right)\right]}{54Qr_0^3\left(r-r_0\right)}+\ldots+\left(\varphi_1^{(0)}+\frac{c_2q^2+4Q^2r_0^4\varphi_2^{(0)}}{3r_0^4Q\left(1+Q^2\right)}+O\left(\omega^2c_1,\omega^2c_2\right)\right)+\ldots,\\
\varphi_2^\text{outer}=&-\frac{\omega^2\left(c_1-c_2\right)}{216r_0^2\left(r-r_0\right)^2}+\ldots+\left[\frac{c_2q^2}{3r_0^5\left(1+Q^2\right)}+\frac{4Q^2}{3r_0\left(1+Q^2\right)}\varphi_2^{(0)}+O\left(\omega^2c_1,\omega^2c_2\right)\right]\left(r-r_0\right)+\ldots,
\end{aligned}
\end{equation}
where, as before, we have only written explicitly the terms which are required for matching, and have neglected the $O\left(\omega^2c_1,\omega^2c_2\right)$ terms as they are subleading in a low $q$ expansion with $\omega\sim q^2$.

To impose incoming boundary conditions at the horizon on the outer solutions (\ref{eq:Hydro1OuterSolnImproved}) and (\ref{eq:Hydro2OuterSolnImproved}), we simply compare the expansions above to those of the ingoing inner solutions in the matching region (\ref{eq:NonZeroTNonHydroInnerMatchingSolutions}). This fixes $c_1$ and $c_2$ to be (at lowest order in an expansion in $\omega$ and $q$)
\begin{equation}
\label{eq:NonZeroTNonHydroIngoingConstantsSolutions}
c_1=c_2=-\frac{6Qr_0^3\mathcal{G}^\text{IR}_{1,T}\left(\omega\right)\left[\varphi_1^{(0)}+\frac{4Q}{3\left(1+Q^2\right)}\varphi_2^{(0)}\right]}{\omega^2+\frac{2q^2}{r_0\left(1+Q^2\right)}\mathcal{G}^\text{IR}_{1,T}\left(\omega\right)}.
\end{equation}
As in the $T=0$ case, $\mathcal{G}^\text{IR}_{2,T}\left(\omega\right)$ contributes to these constants only at a subleading order in the low $\omega$ and $q$ expansion, and is thus outside the range of validity of our approach.

\subsection{Greens functions}

Our final step in computing the retarded Greens functions is to substitute the outer solutions (\ref{eq:Hydro1OuterSolnImproved}) and (\ref{eq:Hydro2OuterSolnImproved}) with ingoing boundary conditions given by (\ref{eq:NonZeroTNonHydroIngoingConstantsSolutions}) into the on-shell action (\ref{eq:OnShellGIAction}). This gives (where the argument $\left(\omega\right)$ is shorthand for $\left(\omega,q\right)$)
\begin{equation}
\begin{aligned}
S=\frac{1}{2\kappa_4^2}&\int_{r\rightarrow\infty}\frac{d\omega dq}{\left(2\pi\right)^2}\frac{3r_0^2q^2\mathcal{G}^\text{IR}_{1,T}\left(\omega\right)}{L^4\left(\omega^2+\frac{2q^2}{r_0\left(1+Q^2\right)}\mathcal{G}^\text{IR}_{1,T}\left(\omega\right)\right)}\Biggl[\varphi_1^{(0)}\left(-\omega\right)\varphi_1^{(0)}\left(\omega\right)\\
&+\frac{16Q^2}{9\left(1+Q^2\right)^2}\varphi_2^{(0)}\left(-\omega\right)\varphi_2^{(0)}\left(\omega\right)+\frac{4Q}{3\left(1+Q^2\right)}\left(\varphi_1^{(0)}\left(-\omega\right)\varphi_2^{(0)}\left(\omega\right)+\varphi_2^{(0)}\left(-\omega\right)\varphi_1^{(0)}\left(\omega\right)\right)\Biggr],
\end{aligned}
\end{equation}
up to contact terms. After using equation (\ref{eq:FieldTheorySourcesInTermsofGIFields}) to express the boundary values of the gauge-invariant fields in terms of the sources of dual field theory operators, we can compute the retarded Greens functions of these dual operators via the usual prescription \cite{Son:2002sd,Kaminski:2009dh}. The results, valid for $T\lesssim\omega\ll\mu$, are identical to those in the hydrodynamic limit given in equation (\ref{eq:HydroGreensFunctionsResult}) and the Ward identities (\ref{eq:WardIdentities}) are automatically satisfied due to our gauge-invariant formulation. These results are valid only at lowest order in $T,\omega$ and $q$ (with $T,\omega\sim q^2$) and are valid only up to contact terms. 

In particular we note that when $T\lesssim\omega\ll\mu$, there is a diffusive mode with dispersion relation (where the ellipses denote terms which are higher order in $q$ and $T$)
\begin{equation}
\label{eq:NonZeroTNonHydroDiffusionDispersionRelationResult}
\omega^2+\frac{2q^2}{r_0\left(1+Q^2\right)}\mathcal{G}^\text{IR}_{1,T}\left(\omega\right)+\ldots=0\;\;\;\;\;\;\;\;\;\longrightarrow\;\;\;\;\;\;\;\;\;\omega=-i\frac{L^2}{3r_0\left(1+Q^2\right)}q^2+\ldots,
\end{equation}
which agrees precisely with the diffusion mode predicted by the hydrodynamic derivative expansion in equations (\ref{eq:ExpectedHydroDiffusionResult}) and (\ref{eq:IntroEtaOverS}). As in the $T=0$ case, it is the dimension of the CFT$_1$ operator dual to the field $\varphi_1$ in the near-horizon Schwarzschild-AdS$_2$ geometry which controls the power of $q$ in the dispersion relation. As we indicated earlier, it is the lack of any $T$-dependence of the CFT$_1$ Greens function $\mathcal{G}^\text{IR}_{1,T}\left(\omega\right)$ which result in there being no terms in the leading order dispersion relation (\ref{eq:NonZeroTNonHydroDiffusionDispersionRelationResult}) which are $q$-independent but $T$-dependent.

\section{Numerical computations of poles and spectral functions}
\label{sec:NumericalResults}

As a check of the validity of our analytic results, we have computed numerically both the poles of the retarded Greens functions and the spectral functions of the operators studied above. This also allows us to reconcile our results with some previous numerical results for the planar RN-AdS$_4$ theory \cite{Edalati:2010hk,Brattan:2010pq}. We have used two different numerical methods to do these calculations.

The simplest method conceptually is to numerically solve the equations of motion (\ref{eq:FinalEoM1}) and (\ref{eq:FinalEoM2}) by integrating out from the horizon of the planar RN-AdS$_4$ spacetime. If one demands ingoing boundary conditions at the starting point of the integration (the horizon), then both the poles of the retarded Greens functions and the spectral functions are easy to determine from the values of the fields at the end point of the integration (the boundary). This numerical procedure is described in detail in \cite{Kaminski:2009dh}.

It is well-known that this method becomes very difficult to implement when $T=0$, as the horizon becomes an irregular singular point of the equations of motion (when $T\ne0$ it is a regular singular point). Leaver's method \cite{Leaver:1990zz} is an alternative way to compute the poles of the Green's functions which does not require integration around a singular point. This method involves making a quasinormal mode ansatz for the fields $\varphi\left(r\right)=\varphi_\text{QNM}\left(r\right)\varphi_{\text{rem.}}\left(r\right)$, where $\varphi_\text{QNM}\left(r\right)$ contains the relevant asymptotic boundary conditions such that $\varphi\left(r\right)$ is ingoing at the horizon and its leading term vanishes at the boundary of the spacetime. After making this ansatz, one then expands the remainder $\varphi_\text{rem.}\left(r\right)$ as a power series (up to order $N_\text{max.}$) around a regular point of the differential equations -- usually taken to be the midpoint between the horizon and the boundary.\footnote{For this method to work, the closest singularities to this regular point should be those at the horizon and the boundary of the spacetime.} One can then expand the equations of motion in a power series up to order $N_\text{max.}$ around the same point to give $N_\text{max.}+1$ algebraic equations for the $N_\text{max.}+1$ coefficients of the power series expansion of $\varphi_\text{rem.}\left(r\right)$. If a non-trivial solution to this set of equations exists, then it implies the existence of a bulk quasinormal mode and hence a pole in the retarded Green's functions of the dual field theory. Note that although this method is most useful at $T=0$ where the method of direct numerical integration is difficult, it can also be applied at non-zero $T$. This numerical procedure is described in more detail in \cite{Denef:2009yy,Edalati:2010hk,Brattan:2010pq}.

\subsection{Zero temperature poles}

As indicated above, we will use Leaver's method to determine the poles of the Greens functions when $T=0$. For the planar RN-AdS$_4$ theory, it is simplest to implement this method using the gauge-invariant Kodama-Ishibashi variables (rather than the variables (\ref{eq:DefinitionofOurGIFields})) for which the equations of motion decouple \cite{Kodama:2003kk}. This was done in detail in \cite{Edalati:2010hk} and we refer the reader to that work for the technical details, as our calculation is identical to that described therein.

Leaver's method indicates the existence of a discrete spectrum of poles with complex frequencies, as well as a very large number of closely-spaced poles along the negative imaginary frequency axis. As described in \cite{Edalati:2010hk}, these purely imaginary `poles' are actually a numerical representation of the fact that there is a branch cut along the negative imaginary $\omega$ axis due to multi-valued terms in the Greens functions (which can be seen in equation (\ref{eq:JottarEtAlResultBranchCutTerms})). If one increases the accuracy of the implementation of Leaver's method by increasing $N_\text{max.}$, the spectrum of these imaginary `poles' becomes denser and denser as they move closer to the origin and in the limit $N_\text{max.}\rightarrow\infty$, this spectrum of `poles' will become a continuous line along the negative $\omega$ axis, which represents the branch cut.

\begin{figure*}
\begin{center}
\includegraphics[scale=0.81]{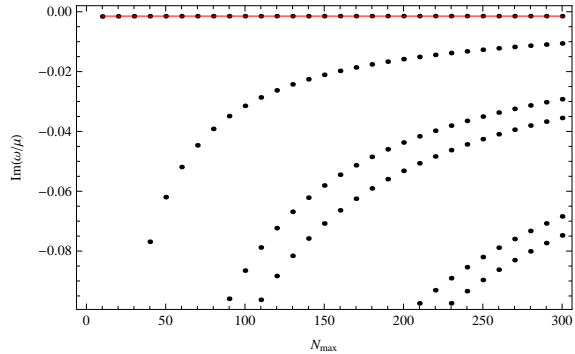}
\caption{The locations of the purely imaginary `poles' nearest the origin as determined by Leaver's method when $q/\mu=0.1$ (shown as dots), as a function of the order of the power series expansion used $N_\text{max.}$. The red line shows the analytic result (\ref{eq:ZeroTResultforDispersionRelation}) for the $T=0$ diffusion mode.}
\label{fig:ZeroTPoleResultsNmaxDependence}
\end{center}
\end{figure*}

In figure \ref{fig:ZeroTPoleResultsNmaxDependence}, we show how the locations of the purely imaginary `poles' are dependent upon the accuracy of the calculation $N_\text{max.}$. 
As expected, most of these `poles' have a location which is not stable -- as one increases $N_\text{max.}$, they move closer together. However, the location of the pole closest to the origin \textit{is} stable as one increases $N_\text{max.}$ and its location agrees precisely with that of the zero temperature diffusion mode predicted by our analytic calculation (\ref{eq:ZeroTResultforDispersionRelation}). We thus conclude that in addition to the branch cut, there is a genuine diffusive pole of the retarded Greens functions with dispersion relation (\ref{eq:ZeroTResultforDispersionRelation}). Our analytic calculation has captured this diffusive pole but not the multi-valued term (which is higher order in $\omega,q$). In this sense, the analytic calculation of the low frequency Greens functions in \cite{Edalati:2010hk} is complementary to ours, as it determines the existence of the multi-valued term, but not of the diffusive pole.

\begin{figure*}
\begin{center}
\includegraphics[scale=0.81]{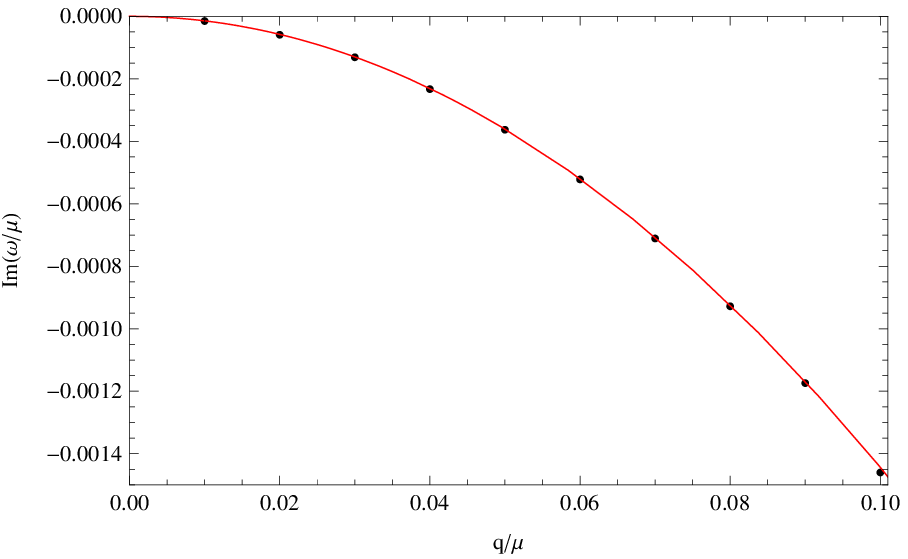}
\includegraphics[scale=0.81]{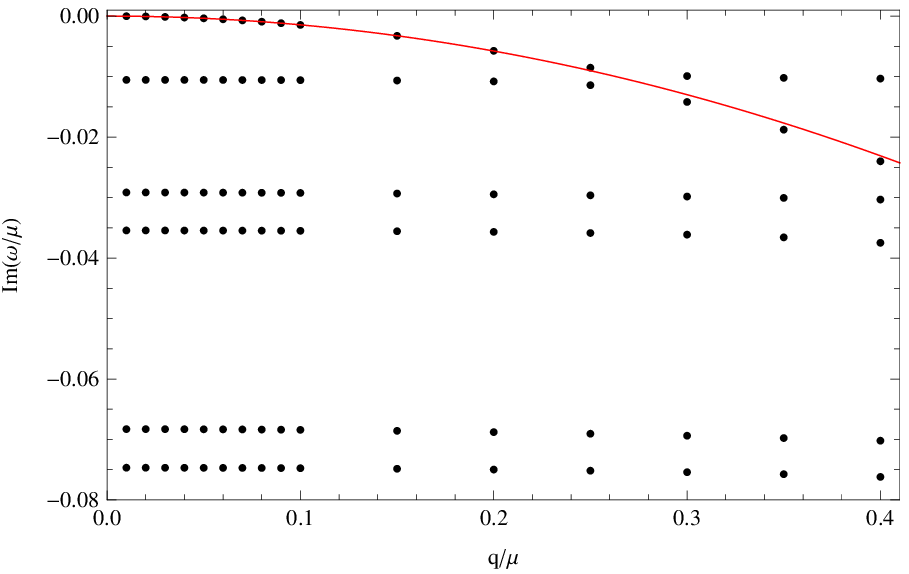}
\caption{Dispersion relations of the purely imaginary `poles' nearest the origin as determined by Leaver's method, at fixed $N_\text{max.}=300$  (shown as dots). The red line shows the analytic result (\ref{eq:ZeroTResultforDispersionRelation}) for the $T=0$ diffusion mode. The data on both plots is the same -- the left hand plot is a magnified version of the near-origin region of the right hand plot.}
\label{fig:ZeroTPoleResultsDispersion}
\end{center}
\end{figure*}

In figure \ref{fig:ZeroTPoleResultsDispersion}, we show the dispersion relations of some of the purely imaginary `poles' at fixed $N_\text{max.}=300$.
The location of the `poles' which are just a representation of the branch cut are approximately $q$-independent. However, the genuine pole has a dispersion relation which is described accurately by our analytic diffusion result (\ref{eq:ZeroTResultforDispersionRelation}) when $q\ll\mu$. Figure \ref{fig:ZeroTPoleResultsDispersion} also shows that the `poles' representing the branch cut extend all the way along the negative imaginary $\omega$ axis to the origin, and do not end at the location of the diffusion pole. Of course, the analytic structure of these poles in the complex $\omega$ plane is not observable. Any observable (e.g. the imaginary part of the Green's function) can be measured only as a function of real $\omega$. The existence of the diffusion pole in the complex $\omega$ plane will strongly influence how these observables depend upon real $\omega$, as can be seen from equation (\ref{eq:HydroGreensFunctionsResult}), whereas the multi-valued terms which produce the branch cut have only a subleading effect.

\subsection{Non-zero temperature poles}

\begin{figure*}
\begin{center}
\includegraphics[scale=0.81]{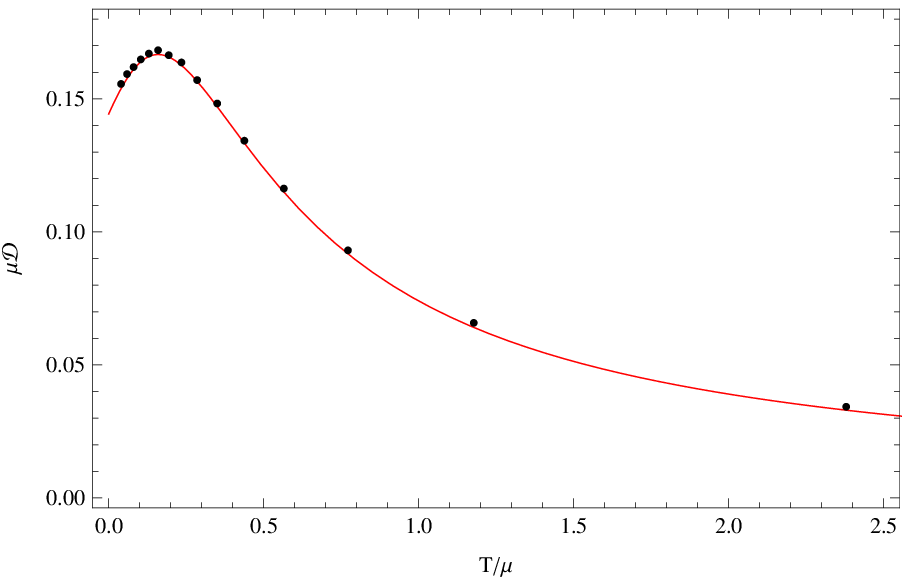}
\includegraphics[scale=0.81]{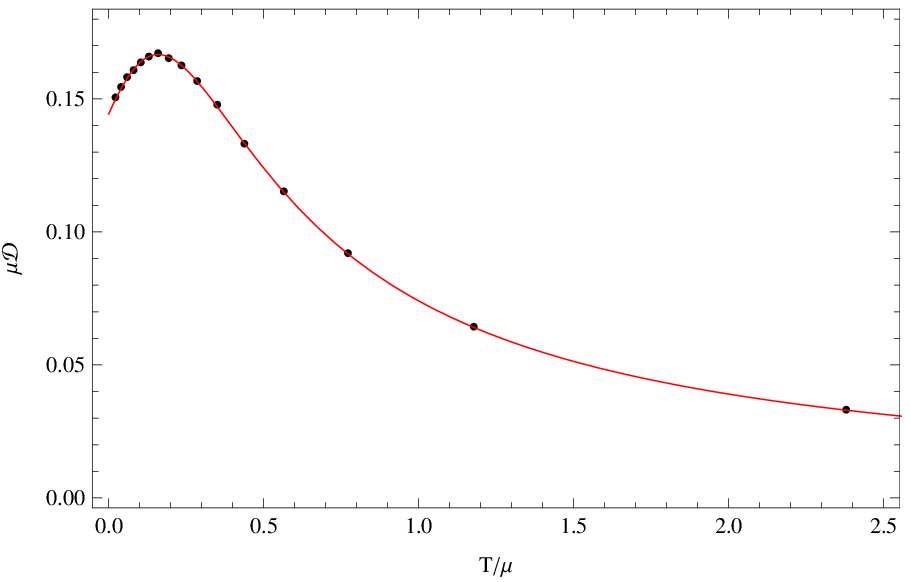}
\caption{The numerical values of the dimensionless diffusion constant $\mu\mathcal{D}$ (shown as dots) as a function of $T/\mu$ as computed by $N_\text{max.}=300$ Leaver's method (left hand plot) and direct numerical integration from the horizon (right hand plot). These were extracted by fitting the function $\omega\left(q\right)=-i\mathcal{D}q^2-i\Delta q^4$ to the numerical location of the pole $\omega\left(q\right)$ at $q/\mu=\left\{0.01,0.05,0.1,0.2,0.3,0.4,0.5\right\}$. The red lines show the hydrodynamic result for $\mu\mathcal{D}$, given by equations (\ref{eq:ExpectedHydroDiffusionResult}) and (\ref{eq:IntroEtaOverS}), which according to our previous analysis should be valid for all $T/\mu$ shown.}
\label{fig:NonZeroTPoleDiffusionConstant}
\end{center}
\end{figure*}

At non-zero temperature, where the branch cut is resolved into a discrete spectrum of poles, it is possible to compute the spectrum of poles using either Leaver's method (as was done in \cite{Edalati:2010hk,Brattan:2010pq}) or direct numerical integration. We computed numerically the dispersion relation of the candidate diffusion mode and fitted this result to a function of the form $\omega=-i\mathcal{D}q^2-i\Delta q^4$, for various values of $T/\mu$. The corresponding numerical values of the diffusion constant $\mathcal{D}$ as a function of $T/\mu$ are shown in figure \ref{fig:NonZeroTPoleDiffusionConstant}.
The results obtained using each numerical method coincide, and are well-described for all temperatures by the hydrodynamic results (\ref{eq:ExpectedHydroDiffusionResult}) and (\ref{eq:IntroEtaOverS}). These results differ from those in figure 9 of \cite{Brattan:2010pq} -- this is because the fitting function used there does not include a $q^4$ term, the effects of which are important towards the upper range of the values of $q/\mu$ that we have studied numerically.

\subsection{Spectral functions}

\begin{figure*}
\begin{center}
\includegraphics[scale=0.81]{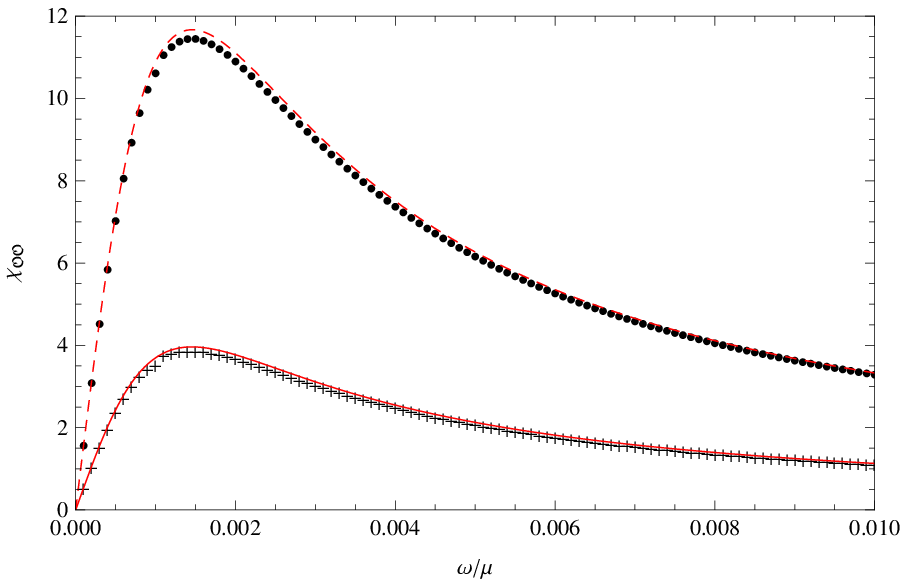}
\includegraphics[scale=0.81]{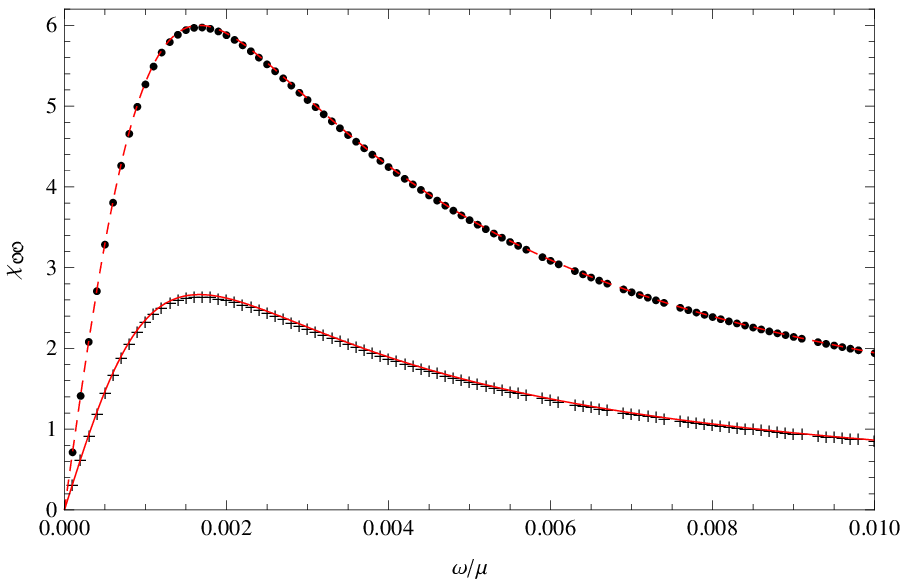}
\caption{Numerical results for the $q/\mu=0.1$ spectral functions $\chi_{T^{ty}T^{ty}}$ (shown as dots) and $\chi_{J^yJ^y}$ (shown as crosses), computed via direct numerical integration from the horizon, for $T/\mu\approx0.005$ (left hand plot) and $T/\mu\approx0.16$ (right hand plot). $\chi_{T^{ty}T^{ty}}$ is plotted in units of $r_0^3/2\kappa_4^2L^4$ and $\chi_{J^yJ^y}$ is plotted in units of $r_0/2\kappa_4^2$. The analytic results (\ref{eq:HydroGreensFunctionsResult}) for $\chi_{T^{ty}T^{ty}}$ (dashed line) and $\chi_{J^yJ^y}$ (solid line) are plotted in red. }
\label{fig:NonZeroTSpectralFunctions}
\end{center}
\end{figure*}

As a further check of our analytic calculations, we have computed numerically the $T\ne0$ spectral functions of the transverse components of $T^{\mu\nu}$ and $J^{\mu}$, which are defined as
\begin{equation}
\chi_{\mathcal{O}\mathcal{O}}\left(\omega,q\right)\equiv-2\text{Im}G^R_{\mathcal{O}\mathcal{O}}\left(\omega,q\right).
\end{equation}
This was done by directly integrating the equations of motion (\ref{eq:FinalEoM1}) and (\ref{eq:FinalEoM2}) from the horizon to the boundary as described in \cite{Kaminski:2009dh}. The resulting spectral functions, at all numerically-accessible temperatures, are well-described by the spectral function formulae computed from our analytic results (\ref{eq:HydroGreensFunctionsResult}). In figure \ref{fig:NonZeroTSpectralFunctions} we show a sample of these results although for conciseness we will not give a detailed exhibition of them here.

\section{Discussion}
\label{sec:Discussion}

In this paper we have studied the two-point functions of the transverse components of $T^{\mu\nu}$ and $J^\mu$ for the field theory dual to the planar RN-AdS$_4$ black hole. Our main result is that a long-lived diffusion mode -- described by the hydrodynamic formulae (\ref{eq:ExpectedHydroDiffusionResult}) and (\ref{eq:IntroEtaOverS}) -- exists outside of the commonly-studied range $\omega\ll T$. Specifically, it exists provided that $\omega\ll T\textit{ or } \mu$. It exists even in the extremal limit $T=0$. 

Recall that hydrodynamics is constructed as a derivative expansion, valid when the length scale over which macroscopic quantities vary is much larger than a characteristic length scale $l_\text{mfp}$. Our results indicate that this length scale is given by
\begin{equation}
l_\text{mfp}=\frac{s}{4\pi\left(\epsilon+P\right)}=\frac{L^2}{3r_0\left(1+Q^2\right)}=\frac{1}{3\mu}\frac{Q}{\left(1+Q^2\right)},
\end{equation}
since with this definition the hydrodynamic diffusion result is valid when $\omega l_\text{mfp},ql_\text{mfp}\ll1$. Note that $Q$ is implicitly a complicated function of $T/\mu$, given in equation (\ref{eq:QasfunctionofTandMu}). In the $\mu=0$ limit, $l_\text{mfp}\sim1/T$ and in the $T=0$ limit, $l_\text{mfp}\sim1/\mu$ (we have neglected numerical factors of order $1$). By definition, this mean free path is equivalent to that suggested in \cite{Bhattacharyya:2007vs} for related global black hole solutions. Our work confirms the assumption therein that the lifetime of the longest-lived mode is proportional to the quantity $s/(\epsilon+P)$ at all temperatures.

A natural extension of the work we have done here is to calculate the low-energy excitations of the longitudinal components of $T^{\mu\nu}$ and $J^\mu$ for this theory outside the usual hydrodynamic range $\omega\ll T$. We expect that the results will be similar in spirit to those presented here in that there will be a sound mode with dispersion relation given by (\ref{eq:ExpectedHydroSoundResult}) and (\ref{eq:IntroEtaOverS}), and a charge diffusion mode, provided that $\omega\ll T\textit{ or }\mu$. There is already numerical evidence to support this claim -- in \cite{Edalati:2010pn} it was shown numerically that there is a $T=0$ sound mode whose real part agrees closely with the hydrodynamic result (\ref{eq:ExpectedHydroSoundResult}) and whose imaginary part differs from the hydrodynamic result by around $10\%$.\footnote{No $T=0$ charge diffusion mode was identified, but this may just be due to the difficulties in numerically disentangling it from the branch cut as described in section \ref{sec:NumericalResults}.} Furthermore, in \cite{Davison:2011uk} the sound dispersion mode was studied at non-zero temperatures. Although not explicitly stated there, both the real and imaginary parts of this mode are consistent with the hydrodynamic results (\ref{eq:ExpectedHydroSoundResult}) and (\ref{eq:IntroEtaOverS}) for all $T$ provided that $\omega\ll\mu$.\footnote{This can be seen by comparing the numerical results for the sound attenuation in figure 11 of \cite{Davison:2011uk} to those for the diffusion constant in figure \ref{fig:NonZeroTPoleDiffusionConstant} of this paper, and noting the relation between these two quantitites predicted by hydrodynamics from equations (\ref{eq:ExpectedHydroDiffusionResult}) and (\ref{eq:ExpectedHydroSoundResult}).} A convincing verification of this expectation will require an analytic derivation of the dispersion relation of the low-energy modes, and we are hopeful that the method we have used here may be suitably modified to provide this.

The low-temperature two-point functions that we have derived are valid only at lowest order in an expansion in small $\omega,q$ and $T$ with $\omega,T\sim q^2$. It would be interesting to try to generalise our low-temperature calculations to include higher order terms in $\omega,q$ and $T$. Firstly, this will allow us to determine more accurately many of the $\omega\rightarrow0$ properties of our theory such as the conductivity and the scaling of the imaginary parts of the Greens functions in this limit. Presently, the calculations of \cite{Edalati:2009bi,Edalati:2010hk} are more accurate in this regard. Perhaps more importantly, this would indicate whether the predictions of hydrodynamics are valid at low temperatures to higher orders in the derivative expansion, or just at the leading order as shown here. Due to the presence of the branch cut, we expect that hydrodynamics will in fact break down at some higher order. In a similar vein, one could investigate the range of validity of non-linear hydrodynamics in this theory by computing higher-point Greens functions. We do not know of an effective theory that incorporates the branch cut in addition to the hydrodynamic collective mode we have found here, and it would clearly be of interest to determine such a theory.

Another obvious question is whether the applicability of hydrodynamics when $T\lesssim\omega\ll\mu$, as shown in this paper, is generic to (non-probe D-brane) holographic field theory states with a large chemical potential. The planar RN-AdS$_4$ solution is in some ways a very special solution, as its entropy density $s$ is non-zero when $T=0$. This feature is intimately linked to its near-horizon AdS$_2\times \mathbb{R}^2$ geometry which is critical for our results since the decay rate of the diffusion mode is controlled by the dimension of a scalar operator in the CFT$_1$ dual to the AdS$_2$ geometry. States with $s=0$ at $T=0$ can be studied by including matter fields in the gravitational theory -- these states typically have a different near-horizon geometry which can, for example, have Lifshitz scaling and/or violate hyperscaling. A particularly interesting example are the near-horizon geometries of \cite{Gubser:2009qt} which are conformal to AdS$_2\times\mathbb{R}^{D-1}$ and have a vanishing zero temperature entropy.

It would be of interest to determine which (if any) of the hydrodynamic results (\ref{eq:ExpectedHydroDiffusionResult}) and (\ref{eq:IntroEtaOverS}) break down at small $T$ for different near-horizon geometries, and if the applicability of these formulae at small $T$ is dependent upon the extent of fractionalisation of the $U(1)$ charge in the dual field theory (which has a strong influence on the near-horizon geometry as reviewed in \cite{Hartnoll:2011fn}). It would be a surprising result if the hydrodynamic results (\ref{eq:ExpectedHydroDiffusionResult}), (\ref{eq:ExpectedHydroSoundResult}) and (\ref{eq:IntroEtaOverS}) are valid at low temperatures irrespective of the dual geometry. This would mean that the leading $q^2$ term in the decay rates of the diffusion and sound modes in `realistic' field theories (i.e.~those with $s=0$ at $T=0$) vanish at $T=0$ and hence these modes would live for an anomalously long time compared to, for example, the well-understood zero temperature sound mode of a Fermi liquid \cite{AbrikosovKhalatnikov,PinesNozieres,BaymPethick}. However, as we have stressed previously, one of the main lessons learned so far from studying strongly-coupled field theories using gauge/gravity duality is that their transport properties can be very different from those one would expect from a quasiparticle-based theory.

\subsection*{Acknowledgements}

We are grateful to Simon Gentle, Juan Jottar and Manuela Kulaxizi for helpful discussions. This work was supported in part by a VIDI innovative research grant from NWO.

\bibliographystyle{JHEP}
\bibliography{RNAdS4Shearv2}

\end{document}